\documentclass[12pt]{iopart}

\pdfoutput=1
\expandafter\let\csname equation*\endcsname\relax
\expandafter\let\csname endequation*\endcsname\relax
\usepackage[english]{babel}
\usepackage{amsmath,
amssymb,amsfonts,latexsym}
\usepackage{graphicx}
\usepackage{color}
\usepackage{subfigure}
\usepackage{afterpage}
\usepackage{booktabs}
\usepackage{blkarray}
\usepackage{cite}
\usepackage{tikz}
\usetikzlibrary{snakes}
\usetikzlibrary{calc}
\usepackage{bm}
\usepackage{braket}
\usepackage{nicefrac}

\definecolor{Blue}{rgb}{0.00, 0.00, 1.00}
\definecolor{Red}{rgb}{1.00, 0.00, 0.00}
\definecolor{labelkey}{cmyk}{.1,.7,0.5,0}

\definecolor{blue(pigment)}{rgb}{0.2, 0.2, 0.6}
\usepackage{mathrsfs}
\usepackage{times}
\usepackage[colorlinks,%bookmarks=false,
citecolor=blue,linkcolor=blue,urlcolor=blue]{hyperref}
\hypersetup{linktocpage}
\usepackage{scalerel}
\usepackage{tensor}
\usepackage{etoolbox}

\makeatletter
\def\@mkboth#1#2{}
\newlength\appendixwidth
\preto\appendix{\addtocontents{toc}{\protect\patchl@section}}
\newcommand{\patchl@section}{%
  \settowidth{\appendixwidth}{\textbf{Appendix }}%
  \addtolength{\appendixwidth}{1.5em}%
  \patchcmd{\l@section}{1.5em}{\appendixwidth}{}{\ddt}%
}
\makeatother

\def\eqref#1{(\ref{#1})}
\usepackage{cancel}

\newcommand{\bea}{\begin{eqnarray}}
\newcommand{\eea}{\end{eqnarray}}

\renewcommand*{\geq}{\geqslant}
\renewcommand*{\leq}{\leqslant}
\newcommand{\I}{\ensuremath{\mathbf{i}}}
\newcommand*{\ddt}[1]{%
  \accentset{\mbox{\bfseries .\hspace{-0.25ex}.}}{#1}} % 2 time derivatives 

\newcommand{\be}{\begin{equation}}
\newcommand{\ee}{\end{equation}}

\newcommand{\beA}{\begin{align}}
\newcommand{\eeA}{\end{align}}

\newcommand{\bV}{\begin{bmatrix}}
\newcommand{\eV}{\end{bmatrix}}

\newcommand{\dd}{\mathrm{d}}
\newcommand{\de}{\partial}

\newcommand{\Ha}{\hat{H}} % Notation for the Hamiltonian

\newcommand{\ssz}{\hat{\sigma}^z} % Notation for spin-chain operators
\newcommand{\ssx}{\hat{\sigma}^x}
\newcommand{\ssy}{\hat{\sigma}^y}
\newcommand{\ssp}{\hat{\sigma}^+}
\newcommand{\ssm}{\hat{\sigma}^-}

\newcommand{\Cc}{\hat{c}}

\newcommand{\rrho}{\hat{\rho}}

\def\pc#1{}

\begin{document}

\title[]{Exact entanglement growth of a one-dimensional hard-core quantum gas during a free expansion}
\author{ Stefano Scopa$^{1}$, Alexandre Krajenbrink$^{1}$, Pasquale Calabrese$^{1,2}$ and J\'er\^ome Dubail$^{3}$}
\address{$^1$ SISSA and INFN, via Bonomea 265, 34136 Trieste, Italy}
\address{$^2$ International  Centre  for  Theoretical  Physics  (ICTP),  I-34151,  Trieste,  Italy}
\address{$^3$ Universit\'e  de  Lorraine,  CNRS,  LPCT,  F-54000  Nancy,  France}
\date{\today}
%\pacs{}
\begin{abstract}
We consider the non-equilibrium dynamics of the entanglement entropy of a one-dimensional quantum gas of hard-core particles, initially confined in a box potential at zero temperature. At $t=0$ the right edge of the box is suddenly released and the system is let free to expand. During this expansion, the initially correlated region propagates with a non-homogeneous profile, leading to the growth of entanglement entropy. This setting is investigated in the hydrodynamic regime, with tools stemming from semi-classical Wigner function approach and with recent developments of quantum fluctuating hydrodynamics. Within this framework, the entanglement entropy can be associated {to} a correlation function of chiral twist-fields of the conformal field theory that lives along the Fermi contour and it can be exactly determined.  Our predictions for the entanglement evolution are found in agreement with {and generalize} previous results in literature based on numerical {calculations} and heuristic arguments. 
\end{abstract}

\newpage
\tableofcontents

\maketitle

\section{Introduction}
The study of the non-equilibrium dynamics of one-dimensional many-body quantum systems covers a large portion of the statistical physics literature (both theoretical and experimental) of the last twenty years. This field has seen the important development of new exotic physics, see e.g.~\cite{Odor2004,Dziarmaga2010,pem-16,Polkovnikov2011,Bertini2021} for modern reviews-- and within it,  the search for analytical results has always been strongly sought. The knowledge of exact results is the cornerstone of clear interpretations of general physical mechanisms involved in non-equilibrium settings and, it can furthermore be used to test new algorithms for the numerical simulations of more involved models, where analytical results are not available.  Out-of-equilibrium statistical mechanics has led over the last decade to important developments such as the concept of of generalized Gibbs ensembles \cite{Rigol2007,Rigol2008,vr-16} and subsequently to the emergence of generalized hydrodynamics (GHD) \cite{Bertini2016,Castro-Alvaredo2016}. Equipped with these tools, it is nowadays possible  to obtain exact results for the out-of-equilibrium evolution of conserved charges and currents \cite{Bertini2016,Castro-Alvaredo2016,Piroli2017,Bulchandani2017,DeLuca2017,Doyon2017,DeLuca2017b,Bulchandani2018,Doyon2018,Bastianello2019,Mestyan2019, Schemmer2019,Bastianello2020a,Ruggiero2020,Malvania2020,Bouchoule2020a} as well as correlation functions \cite{Doyon2018,Perfetto2020,Moller2020,d-ls} and transport properties (such as Drude weights \cite{Ilievski2017,Bertini2021}) of a large number of integrable models, see also \cite{Bertini2021,Alba2021,d-ls} as recent reviews. The transport properties of one-dimensional quantum systems has been also investigated using non-homogeneous (1+1) dimensional conformal field-theories, see \cite{sc-08,Dubail2017,Brun2017,Brun2018,Langmann2019,Moosavi2019,Ruggiero2019,Scopa2020,Collura2020,Gluza2020}.
In non-interacting models, another recent effort was conducted to classify universal results with the use of random matrix theory\cite{Dean2019,Dean2018,Smith2020,Gautie2021,DeBruyne2021}.  
{Distinctive non-equilibrium features are usually and simply encoded into the entanglement entropy dynamics, which however is difficult 
to characterize from ab-initio calculations}, due to its computational complexity. 

{The} case of {\it global} quantum quenches \cite{Calabrese2006,Calabrese2007a}, 
where the out-of-equilibrium {dynamics is initiated} by a sudden variation of one or more  Hamiltonian parameters {homogeneously}, 
{can be studied in integrable models} using the quench-action formalism \cite{Caux2013,Brockmann2014,Ilievski2015,Caux2016}, 
through which  time-dependent out-of-equilibrium quantities are derived under the assumption that relevant states are found in the vicinity of the saddle point of a certain functional. 
In {this} global quench setting, the {\it linear} growth of the entanglement entropy is well captured by the quasi-particle picture \cite{Calabrese2005,Alba2017,ac-18,c-20}, where one assumes that quasi-particles emitted from different points are unentangled while pairs of quasi-particles emitted from the same point are entangled and, as they move far apart, they are responsible for the spreading entanglement and correlations throughout the entire system. This construction has proved to be fruitful to the point that it has been extended to non-homogeneous settings,  by incorporating GHD in its most recent formulation \cite{bfpc-18,Alba2019,a-19}. The situation is different for {\it local} quenches, where a slower -- and typically {\it logarithmic} -- growth of entanglement is expected, see e.g.~\cite{Calabrese2007,Eisler2008,Eisler2009,Igloi2009,Stephan2011,cc-13,Calabrese2016}.

{Halfway} between these two scenarios, a peculiar class of quantum quenches is that of bi-partitioning protocols see e.g.~\cite{Antal1999,Karevski2002,Vicari2012,Alba2014,chl-08,Allegra2016,Castro-Alvaredo2016,Dubail2017,Bertini2016,DeLuca2017,DeLuca2017b,Piroli2017,Bertini2018,Gruber2019,Collura2020}, where a logarithmic growth of entanglement \cite{Vicari2012,Dubail2017,Alba2014,Gruber2019,Collura2020} is associated with the integrability of the model. In this context, a remarkable example is the time-evolution of a domain-wall state $\ket{\uparrow\dots\uparrow\downarrow\dots\downarrow}$ of a spin chain model or equivalently a fermionic chain filled entirely on one of its halves, whose properties (conserved charges and entanglement respectively) have been fully characterized in both the non-interacting \cite{Antal1999,Karevski2002,Lancaster2010,Allegra2016,Dubail2017} and interacting \cite{DeLuca2017,Collura2020} cases. A similar non-equilibrium setting consists in joining a spin chain having a state $\ket{\psi}$ with a non-saturated value of magnetization {(typically the ground state)}, to another spin chain in a ferromagnetic reference state, e.g. $\ket{\psi}\otimes\ket{\downarrow\dots\downarrow}$
{(in some literature this protocol is known as a geometric quench \cite{mpc-10,Alba2014,Gruber2019}).
This setup has been studied intensively in the past} and the exact expressions for the conserved charges and current profiles have been found 
for a spin-1/2 ${\rm xxz}$ model both in the free \cite{Antal1999,Antal2008} and interacting \cite{Gruber2019} cases. 
{Very recently, it was also shown that initial quantum correlations can suppress particle transport \cite{Jin2021}.}
On the other hand, {so far the exact evolution of the entanglement entropy has been only conjectured} on the basis of heuristic arguments and numerical results \cite{Vicari2012,Alba2014,Gruber2019}.

In this work, we analytically investigate the evolution of the entanglement entropy of the initial state $\ket{\psi}\otimes \ket{\downarrow\dots\downarrow}$, focusing on the 
non-interacting spin-$1/2$ {\rm xx} chain. Throughout the rest of the paper, we shall follow an equivalent interpretation of this setup in terms of a quantum gas made of hard-core particles (free spinless Fermi gas or, equivalently, hard-core bosons) at zero temperature. In particular, the gas is initially confined in the region $[-L,0]$ by a box potential with infinitely high edges. At $t=0$, the right wall is switched off and the gas is let free to expand to the right hand side of the system.
{This is nothing but a quantum version of the textbook Joule expansion}.

The quantum gas is studied at hydrodynamic scales where the post-quench evolution of the model is suitable described in terms of the Wigner function. The latter satisfies a simple transport equation that provides a very-intuitive interpretation of the non-equilibrium dynamics in phase-space. This semi-classical picture is then complemented with quantum fluctuations of the Fermi contour \cite{Ruggiero2019,Ruggiero2020}, described in terms of a Luttinger-liquid model \cite{Cazalilla2004,Giamarchi2007}. 
The computation of the entanglement entropy requires instead the use of conformal-field theory (CFT) and a regularization of the initial state in the phase-space, 
whose details are provided in the main text. 
Within this framework, we find an analytical expression for the entanglement entropy evolution which {does not depend on the used regularization}. 
Moreover, our formula predicts an asymptotic growth of the entanglement entropy at position $x=0$ as $S_1(0,t)\sim\frac{1}{4}\log t$, in contrast with the typical $\frac{1}{6}\log t$ growth which is found for an initial domain-wall state \cite{Dubail2017,Collura2020}.
The {study of the} interacting case goes beyond the scope of this work and will be therefore addressed in a subsequent publication \cite{next-pub}.

\paragraph{Outline.}
 The paper is organized as follows. In Section~\ref{sec:model}, we introduce the lattice model and  set up the {considered quench protocol}. 
 Moreover, we summarize the previous results in literature for the entanglement entropy {and compare them} with the exact prediction obtained in this work. Section~\ref{sec:hydro-limit} is dedicated to the hydrodynamic limit of the problem. After defining the model at hydrodynamic scales, we proceed by introducing the Wigner function of the coarse-grained model and with it, we establish a semi-classical hydrodynamic description of the time evolution in phase-space. Afterwards, we incorporate quantum fluctuations on top of the classical hydrodynamic background by introducing density fluctuating fields, whose dynamics is governed by a Luttinger-liquid theory. 
In Section~\ref{sec:entanglement}, we explain the strategy of the computation of the entanglement entropy using a CFT approach. Here one can find the details of the derivation, including the regularization, in the CFT sense, of the initial problem.
Several numerical {exact calculations} have been performed to test and complement our findings.  Finally, Section~\ref{sec:conclusion} contains our conclusions as well as a few outlooks. Technical details on the {numerics} and further analysis on the entanglement evolution can be found in two appendices, \ref{app:numerics} and \ref{app:param-Jerome}.

\section{Setup and main result \label{sec:model}}

\subsection{The model: free expansion of {a} lattice hard-core gas}
We consider a semi-infinite one-dimensional lattice $j\in[-L,+\infty]$ loaded with hard-core particles with nearest-neighbor hopping and coupled to a potential $V$, described by the Hamiltonian
\be\label{free-gas}
\Ha= -\frac{1}{2}\sum_{j=-L}^{+\infty} \left(\Cc_j^\dagger \Cc_{j+1} + \Cc_{j+1}^\dagger \Cc_j \right)  \, + \, \sum_{j=-L}^{+\infty} V_j\ \Cc_j^\dagger \Cc_{j} .
\ee 
Here the creation and annihilation operators $\Cc^\dagger_j$ and $\Cc_j$ are lattice fermion operators that satisfy the canonical anti-commutation rule, $\{ \Cc_i , \Cc_j^\dagger  \} = \delta_{ij}$. It is well known that this model is equivalent to hard-core bosons, or, also, to the spin-$1/2$ ${\rm xx}$ chain
\be\label{xx-model}
\Ha= - \frac{1}{4} \sum_{j=-L}^{+\infty} \left(\ssx_j \ssx_{j+1} + \ssy_j\ssy_{j+1}\right) \, + \, \frac{1}{2}\sum_{j=-L}^{+\infty} V_j \ssz_j \, + \, {\rm constant}
\ee
where $\hat\sigma_j^{a}$, $a=x,y,z$, are spin-$1/2$ operators acting at site $j$. Both forms of the Hamiltonian are related through the Jordan-Wigner transformation \cite{Jordan1928}
\be
\Cc_j^\dagger= \exp\left(\I \pi \sum_{i<j} \ssp_i\ssm_i\right) \ssp_j , 
\ee
where $\hat\sigma^\pm_j=(\ssx_j\pm \I \ssy_j)/2$.

\begin{figure}[t]
\centering
	\includegraphics[width=0.5\textwidth]{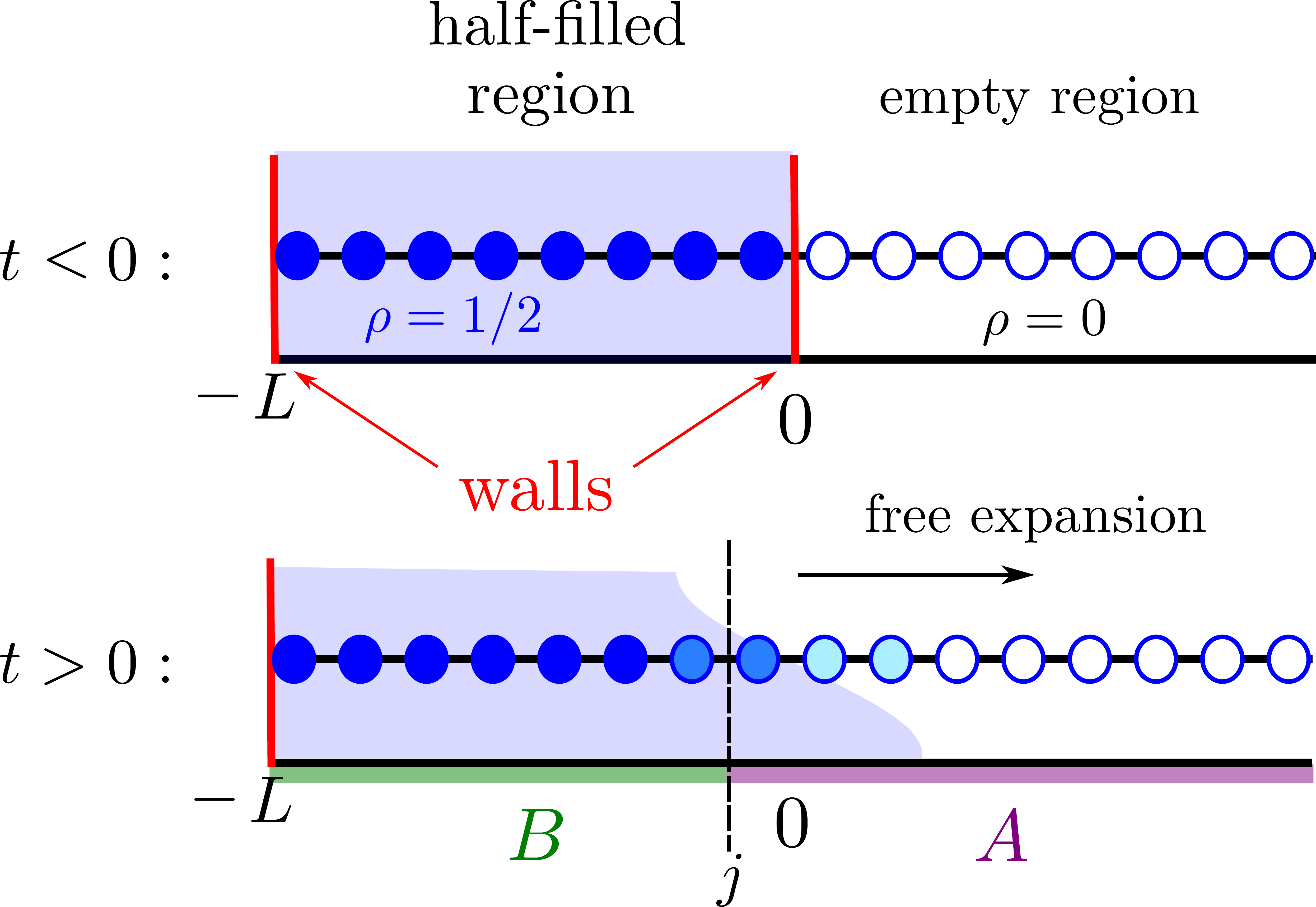} \includegraphics[width=0.45\textwidth]{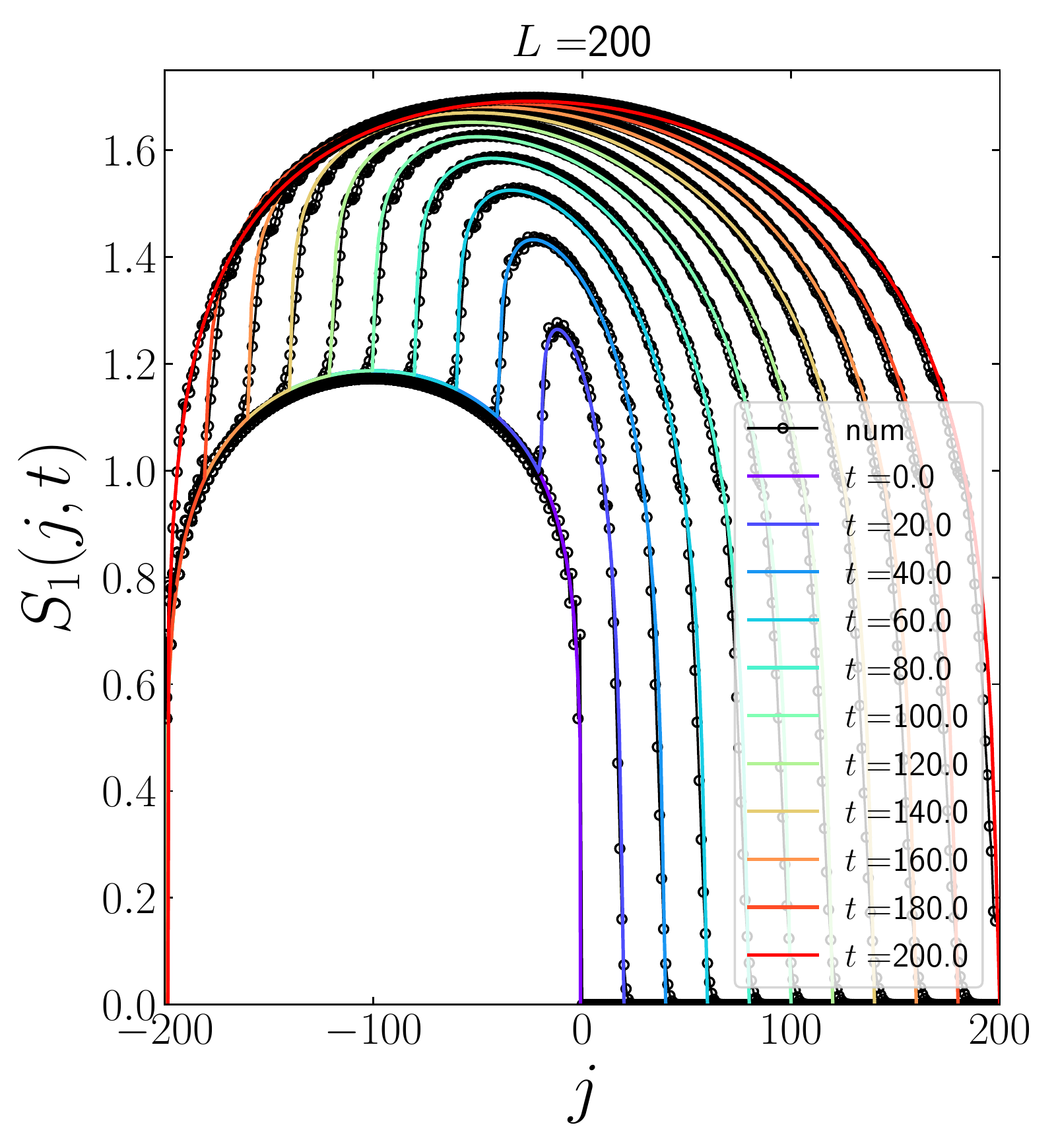}\\
{\bf(a)} \hspace{5cm} {\bf (b)}\\
\caption{{\bf (a)} {\bf Illustration of the setup}: free expansion of a lattice gas at zero temperature, initially confined in the box $[-L,0]$. At $t=0$ the infinite wall at $j=0$ is removed, and for $t>0$ the gas freely expands to the right. During the expansion, the gas is inhomogeneous, with a density that depends on position $j$ and $t$, depicted with a light-blue area in the figure. The exact density profile was first computed in Refs.~\cite{Antal1999,Antal2008}. 
{\bf (b)} The main result of this paper is the exact evolution of the entanglement entropy $S_1(j,t)$ of the subsystem $A = [j, + \infty]$, obtained by tracing 
out the subsystem $B = [-L,j-1]$. 
The black circles correspond to the exact {numerical results} for a lattice of size $2L$ (see text), while the colored curves are the analytic formula~\eqref{solution}.
For simplicity, we restrict to times $t \leq L$ (see text). 
}\label{fig:summary}
\end{figure}

We assume that the gas is initially in the ground state of the Hamiltonian \eqref{free-gas} with the infinite-wall potential
\begin{equation}\label{potential}
	t<0: \qquad  V_j \, = \,  \left\{ \begin{array}{rcl}
		0  &{\rm if}& j \in[-L,-1] \\
		+\infty  &{\rm if} & j\geq 0 .
	\end{array} \right.
\end{equation}
In the absence of an additional chemical potential, the ground state {(in the grand canonical ensemble where the particle number is not a priori fixed)}
contains exactly $L/2$ particles (we assume $L$ is even). This can be easily seen by diagonalizing the Hamiltonian~\eqref{free-gas} with the potential \eqref{potential}, 
\be
\Ha_{t<0} =- \sum_k   \cos(k)\;  \hat\eta_k^\dagger \hat\eta_k,
\ee 
with Fourier modes of momentum $k= \pi q/(L+1)$, $q=1,\dots, L$, given as
\be
\hat\eta_k^\dagger= \sqrt{\frac{2}{L+1}} \sum_{j=-L}^{-1}  \sin \left( k j \right) \Cc_j^\dagger,\qquad  \{ \hat{\eta}_k^\dagger, \hat{\eta}_{k'} \}=\delta_{kk'}.
\ee
Indeed, the single-particle energy $-\cos(k)$ is negative for $q=1, \dots,L/2$, so the ground state is obtained by acting on the fermion vacuum $ \left| 0 \right>$ with those single-particle creation modes, $\hat{\eta}_1^\dagger  \hat{\eta}_2^\dagger \dots \hat{\eta}_{\nicefrac{L}{2}}^\dagger  \left| 0 \right>$.
Then, at times $t>0$, the infinite wall at the origin is switched off,
\begin{equation}
	t>0: \qquad  V_j \, = \, 0,
\end{equation}
and the gas expands freely to the right, see Fig.~\ref{fig:summary}{\bf(a)}. 

In the hydrodynamic limit $L\to \infty$,  $t \rightarrow \infty$, $ j  \rightarrow \infty $ with $t\leq L$ and $j/t$ fixed (see Sec.~\ref{sec:hydro-limit}), the density profile is modified in the region $-t\leq j \leq t$ according to~\cite{Antal1999,Antal2008}
\begin{equation}\label{dens-intro}
	\rho(j,t) \, = \, \frac{1}{2\pi} {\rm arccos} \frac{j}{t},
\end{equation}
at times $0<t \leq L$. Thus, as the gas expands, the initially correlated region propagates towards the right hand side with a non-homogeneous profile, {spreading the 
entanglement on the right side}. In particular, in this paper we will focus on the growth of the $\alpha$-R\'enyi entropy of the reduced density matrix of the subsytem $A= [j,+\infty]$,
\be
S_\alpha(j,t)= \frac{1}{1-\alpha} \log  \tr \ (\rrho_A(t) )^\alpha
\ee
%from the left hand side of the chain towards the right vacuum 
and to its limit $\alpha \to 1$, where it reduces to the Von Neumann entanglement entropy 
\be
S_1(j,t)= - \tr \rrho_A(t) \log \rrho_A(t).
\ee
{The large-scale properties of the Von Neumann and R\'enyi entropy will be accessed exploiting} the hydrodynamic property of the quantum gas \eqref{free-gas}.

%%%%%%%%%%%%%%%%%%%%%%%%%%%%%%%
%%%%%%%%%%%%%%%%%%%%%%%%%%%%%%%

\subsection{Previous results in the literature}
We briefly {recap} the previous results in {the} literature for the entanglement evolution in the {free gas expansion} setup discussed in the previous section. 
To our best knowledge, a first analysis {was performed} in Ref.~\cite{Alba2014}, where a conjectured formula for the asymptotic growth 
{of the half system entanglement (i.e.  $j=0$) matched very well the numerical data}. 
This conjecture has been then extended in Ref.~\cite{Gruber2019} {to} the following ansatz for the entire entanglement profile
\be\label{Eisler-ansatz}
S^{\rm ansatz}_1(j,t)=\frac{1}{6}\log\left(\frac{L}{\pi} \sin\left(\frac{\pi(t-j)}{2L}\right)\right) +\frac{1}{12}\log\left((t-j)(1-\frac{j^2}{t^2})\right)+\kappa
\ee
{expected to hold for any} $|j|< t$. {Here} $\kappa$ is an additive constant treated as a fitting parameter. 
{For $j=0$, Eq. \eqref{Eisler-ansatz} reduces to the one of} Ref.~\cite{Alba2014}. 
As argued below, Eq.~\eqref{Eisler-ansatz} provides a good {description} of the entanglement entropy, but it is different from the exact result obtained in this work.

\subsection{Main result of this paper}

The main result of this paper is an exact asymptotic formula for the entanglement entropy of the subsystem $A=[j,+\infty]$  at time $t$ in the hydrodynamic limit, 
\be\label{solution}
\begin{split}
&S_1(j,t)=\\
&\begin{cases} \frac{1}{6}\log
\left(\frac{L}{2\pi}{\sqrt{\left|\frac{j}{t}-t(1-\frac{j^2}{t^2})\right|}} \ \left|\sqrt{1+\sqrt{1-\frac{j^2}{t^2}}}-{\rm sign}(j)\sqrt{1-\sqrt{1-\frac{j^2}{t^2}}}\right| \left|\sin(\frac{\pi(j-t)}{2L}) \right| \right)+\Upsilon \\[4pt]
\hspace{13.3cm} \text{if $|j|< t$};\\[8pt]
\frac{1}{6}\log\left({\frac{L}{\pi}}\left|\sin\frac{\pi j}{L}\right|\right) +\Upsilon \qquad \text{if $j\leq-t$};\\[8pt]
0, \qquad \text{otherwise}
\end{cases}
\end{split}
\ee
where $\Upsilon\simeq 0.49502$ is a known non-universal constant\cite{Jin2004}. %The asymptotic entanglement profile in Eq. \eqref{solution} is discontinuous at $j = -t$. In any microscopic model or in a regularized field-theory, the discontinuity is smoothed on the scale of the cutoff, making the profile continuous. This phenomenon will be explicitly seen for the regularization that we discuss in Sec.~\ref{sec:solution}.
 In Figure~\ref{fig:summary}{\bf(b)}, we show the result in Eq.~\eqref{solution} compared 
with exact  {numerical results}. Importantly, taking $j=0$ in our formula, we find that the entanglement entropy grows as
\begin{equation}\label{asy-prediction}
	S_1(0,t) \, = \, \frac{1}{4} \log t , \qquad  {\rm for } \quad L \gg t,
\end{equation}
in agreement with the numerical observation of Refs.~\cite{Alba2014,Gruber2019}. We stress that, although logarithmic dependence on time or system size of the entanglement entropy is a ubiquitous phenomenon in one-dimensional quantum critical systems, the coefficient $\frac{1}{4}$ is very unusual in a system with central charge $c=1$. To our knowledge, the calculation we present in this paper is the first analytical derivation of that unusual prefactor $\frac{1}{4}$.  

\begin{figure}[t]
\centering
\includegraphics[width=0.65\textwidth]{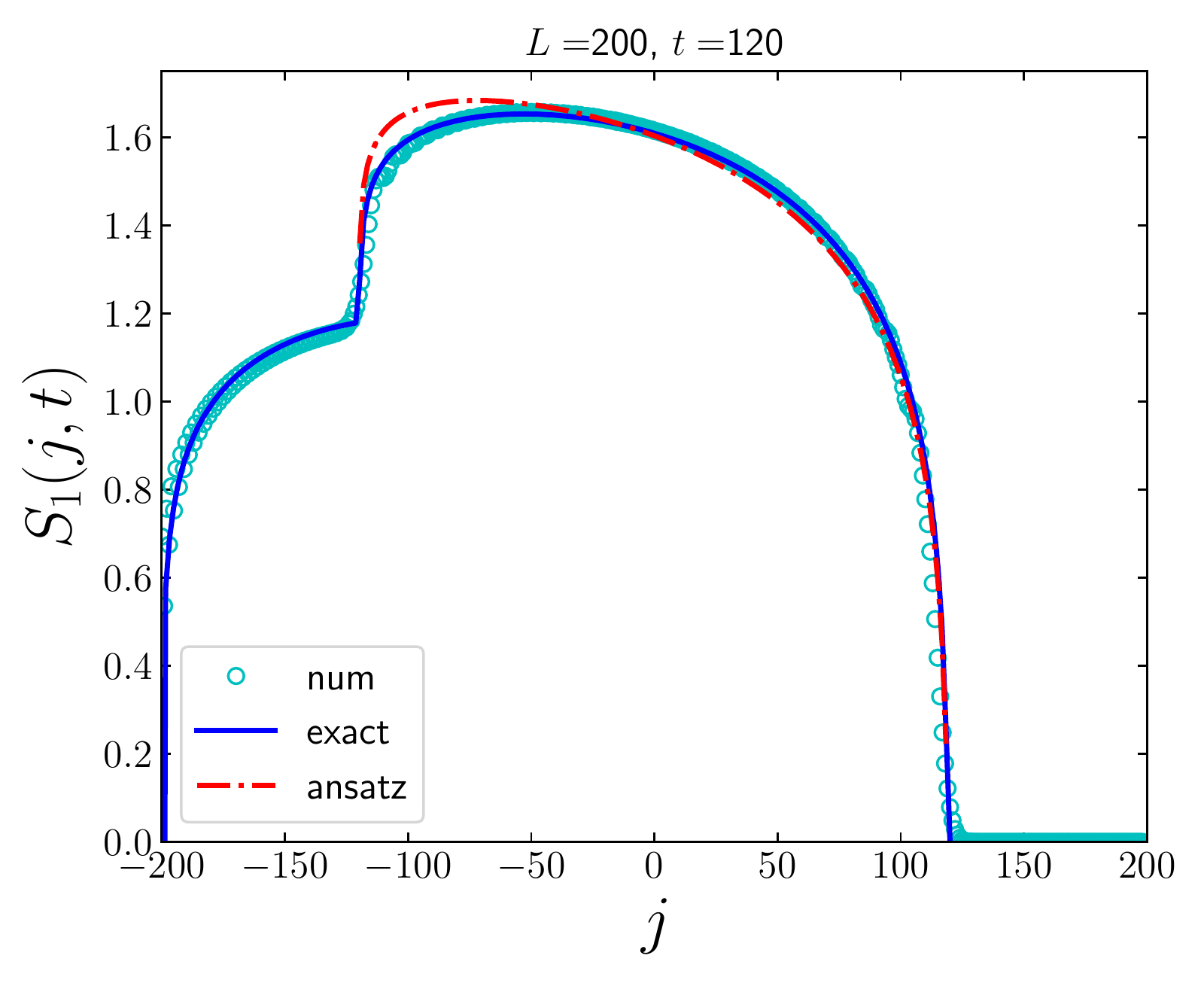}
\caption{\label{fig:comparison} Comparison between our exact result in Eq.~\eqref{solution} ({full line}) and the result of Ref.~\cite{Gruber2019} in Eq.~\eqref{Eisler-ansatz} ({dot-dashed line}), where $\kappa$ has been fixed with a best-fit of \eqref{Eisler-ansatz}. 
Although the two curves are numerically very close, our result \eqref{solution} shows a better agreement with the numerics {(symbols)} for any value of $j$.}
\end{figure}

Our exact analytical formula \eqref{solution} also shows that, beyond the special case $j=0$, the result of Ref.~\cite{Gruber2019} for the entanglement profile is not correct, although, numerically, the ansatz in Eq.~\eqref{Eisler-ansatz} is quite close to \eqref{solution}, see Fig.~\ref{fig:comparison}.
%%%%%%%%%%%%%%%%%%%%%%%%%%%%
%%%%%%%%%%%%%%%%%%%%%%%%%%%%

\section{Classical and quantum fluctuating hydrodynamic description}\label{sec:hydro-limit}

We consider a continuous hydrodynamic description of the lattice Hamiltonian \eqref{free-gas}. Although it is possible to obtain exact results for several quantities in free Fermi models within their lattice formulation, the hydrodynamic approach allows {us} to avoid involved calculations and gives an easy access to asymptotically exact results. In addition, as we discuss in Sec.~\ref{sec:entanglement}, it {leads to} exact asymptotic results for the entanglement entropy evolution, whose lattice derivation is generically very demanding and, for {our} problem, it is out-of-reach with known techniques {(although results for the stationary states in similar settings
may be worked out, see e.g. \cite{fg-21})}. Therefore, throughout the rest of this work we consider the hydrodynamic limit of the setup of Sec.~\ref{sec:model}, keeping its lattice formulation for the numerical test of our results.

\subsection{Continuum limit and regularization of the problem}

We first divide the chain into equally-spaced intervals of size $\Delta x=M\delta$, each of them containing a large number $M\gg 1$ of lattice sites, where $\delta$ is the lattice spacing. The continuum limit consists in having the double limit $\delta\to 0$, $\Delta x\to 0$ keeping the ratio $\Delta x/\delta=M$ fixed. The lattice site $j$ is subsequently replaced by a continuous variable $x=j\delta\in\mathbb{R}$ and the Hamiltonian \eqref{free-gas} can be further written as \cite{Wendenbaum2013}
\be\label{xx-confined-continuum}
\Ha=\int_{-L}^{+\infty} \dd x \ \int_{0}^{\Delta x} \frac{\dd y}{\delta \ \Delta x} \ \left[-\frac{1}{2}\left(\Cc^\dagger_{x+y} \ \Cc_{x+y-\delta} +{\rm h.c.} \right) + \Cc_{x+y}^\dagger \ V(x) \ \Cc_{x+y} \right],
\ee
with continuous fermionic fields $\Cc^\dagger_x\equiv\Cc^\dagger_{j\delta}=\Cc^\dagger_j$, $(\Cc^\dagger_x)^\dagger=\Cc_x$. 
{The hydrodynamic limit above corresponds} to a description of the model over mesoscopic scales such that $L\gg \Delta x \gg \delta$. Moreover, in writing Eq.~\eqref{xx-confined-continuum}, we assumed that the potential is varying slowly enough to be considered constant in each cell of size $\Delta x$, i.e. $V(x+y)\simeq V(x)$ for each $y\in[x,x+\Delta x]$. This in turns requires a regularization of the infinite-wall potential \eqref{potential} that we replace with
\be\label{exp-potential}
V(x)= \exp(\beta x), \qquad \beta>0 .
\ee
We further consider the double scaling limit where $\beta\to\infty$, $\Delta x\to 0$ keeping  $\beta \Delta x \ll 1$ in such a way that the potential reproduces a infinite-wall confinement at large scales but it can be still considered to be constant inside each coarse-grained interval $\Delta x$. Such assumption is generically referred to as {\it scale separation hypothesis} \cite{Wendenbaum2013,Allegra2016,Dubail2017,Brun2017,Brun2018,Ruggiero2019,Bastianello2020,Scopa2020,Ruggiero2020}.  It is then easy to see that 
{the limit}  $\beta \to \infty$ corresponds to 
\be\label{beta-limit}
\beta \to \frac{1}{\Delta x} \sim \frac{1}{\delta}\ ,
\ee
since $\delta$ sets the UV scale of the problem.  With this choice, the ground state of the {trapped} gas in Eq.~\eqref{xx-confined-continuum} reproduces the ground state of the lattice Hamiltonian \eqref{free-gas} in the appropriate scaling limit. At time $t=0$, the trap $V$ is suddenly removed and the system evolves according to the Hamiltonian dynamics.

{An} equivalent hydrodynamic description of the model can be achieved by considering the scaling limit $L\to\infty$, $j\to\infty$, $t\to \infty$ at fixed ratio $j/t$. From this perspective, all the quantities are measured in units of $\delta$ and the parameter $\beta$ in Eq.~\eqref{beta-limit} is thought as a constant $\sim{\cal O}(1)$. The infinite-wall limit of the model is then recovered asymptotically for $L\to \infty$. Although the two hydrodynamic limits of the Hamiltonian \eqref{free-gas} are equivalent, we will make use of the former, treating $\beta$ as a tunable parameter at large (but fixed) $L$ for the rest of the work, while, we keep the latter only for a direct comparison with numerical {calculations}.
 Moreover, for a better exposition, we write the rescaled time variable {as} $\tau\equiv t \delta$, keeping the notation $t$ for the time measured in $\delta$ units.

\subsection{Classical hydrodynamic description}\label{sec:hydro-classical}
At hydrodynamic scales, the gas in each coarse-grained point $x$ is assumed to be the eigenstate of a system in a periodic box of size $\Delta x$. It follows that the Hamiltonian \eqref{xx-confined-continuum} can be diagonalized in Fourier space
\be
\Ha=\int_{-L}^{\infty} \dd x \; \int_{-\pi}^\pi \frac{\dd k}{2\pi} \left[V(x)- \cos k\right] \ \hat\eta^\dagger_{k,x} \ \hat\eta_{k,x},
\ee
with Fourier modes 
\be
\Cc^\dagger_{x+y}=\int_{-\pi}^\pi \frac{\dd k}{2\pi} \ e^{\I k y} \;\hat\eta^\dagger_{k,x} \qquad \hat\eta_{k,x}=(\hat\eta^\dagger_{k,x})^\dagger.
\ee
Its ground state is populated by particles of momentum $|k|\leq k_F(x)$, where $k_F$ is the local Fermi momentum obtained from the semi-classical energy
\be
\varepsilon(x,k)=-\cos k + V(x)
\ee
as
\be\label{kF-initial}
k_F(x)=\arccos V(x).
\ee 
The left hand side of the system $x\in [-L,0]$ is initially populated by modes $k$ such that $k\leq k_F(x)$,  while the right hand side $x\in (0,+\infty)$ is characterized by the absence of particles. At times $\tau\geq 0$, each mode $k$ propagates ballistically with a constant velocity $v(k)=\sin k$, due to the non-interacting nature of the system. From this simple picture one can conclude that the modes $k$ found around a spatial point $x$ at time $\tau$ are those emitted from a position $x_0$ such that  the equation of motion 
\be\label{eq-of-motion}
x=x_0+ v(k) \ \tau
\ee
is satisfied. Moreover, the particles spreading is bounded by the propagation velocity of the fastest modes $k=\pm \pi/2$, {$|v(\pm\pi/2)|=1$, which 
define the {\it light-cone region} $|x/\tau|\leq 1 $}, outside of which the system keeps its initial configuration.  Close to the light-cone, there is a boundary layer of width $x/\tau \pm 1\sim \tau^{-2/3}$ which allows the matching of the physics of the bulk of the light-cone with the one outside of it \cite{Eisler2013,Moriya2019, Jin2021}. We mention that within our hydrodynamic approach it is possible to establish the correct behavior also in this layer, see e.g.~\cite{Bettelheim2011,Bettelheim2012,Allegra2016}, but this analysis goes beyond the scope of this paper.

\begin{figure}[t!]
\centering
\includegraphics[width=0.7\textwidth]{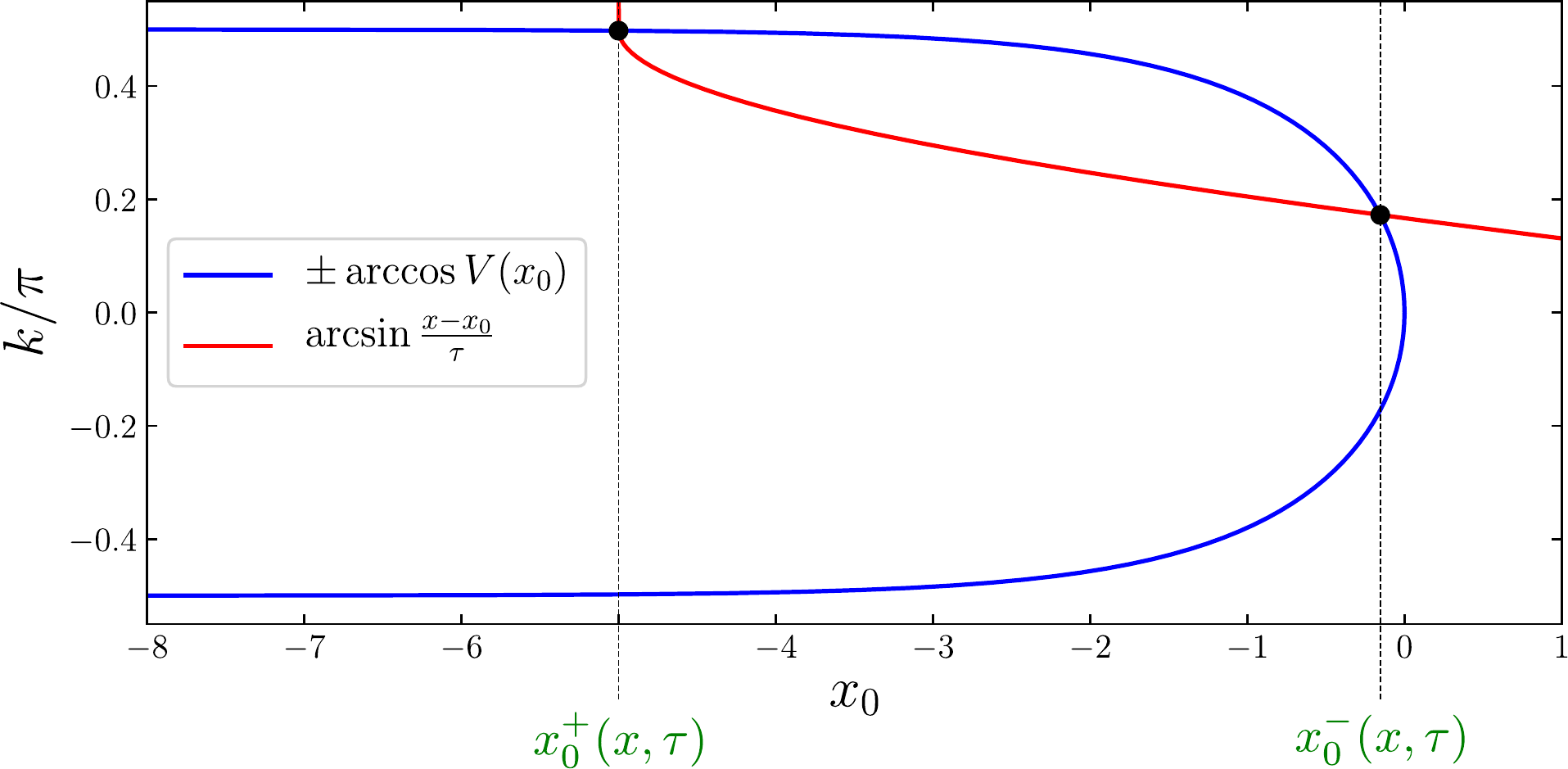}
\caption{\label{fig:roots} Graphical solution of Eq.~\eqref{self-const} for a given value of $x$, $\tau$. In the figure, the two roots $x_0^\pm(x,\tau)$ (dashed vertical lines) has been numerically computed for $x=5$, $\tau=10$ and $\beta=1$.}
\end{figure}

Hence, from Eq.~\eqref{eq-of-motion} one finds that the allowed modes $k$ at position $x$ and time $\tau$ are those belonging to the interval  $[k_F^-, k_F^+]$, where the Fermi points $k_F^\pm$ correspond to particles initially emitted at positions $x_0^\pm$ such that
\be\label{kF-def}
k_F^\pm(x,\tau) =\arcsin\frac{x-x_0^\pm}{\tau}.
\ee
Since the momentum of the excitations is conserved during the time evolution, we can further impose that $|k_F^\pm|=\arccos V(x_0^\pm)$, obtaining
\be\label{self-const}
\frac{(x-x_0^\pm)^2}{\tau^2}= 1-V(x_0^\pm)^2=1-\exp\left(2\beta x_0^\pm\right)
\ee 
which can be solved numerically for the initial position $x_0^\pm$ of the Fermi points $k_F^\pm$, see Fig.~\ref{fig:roots}.  Notice that equation~\eqref{self-const} has always two solutions for $x\leq \tau$ and zero solution when $x>\tau$. 

%%%%%%%%%%%%%%%%%%%%%%%%%%
%%%%%%%%%%%%%%%%%%%%%%%%%%

By summing up each individual contribution coming from a filled mode at $x/\tau$, one can obtain several quantities {of interest}, as for instance the particle density profile during the gas expansion
\be\label{dens-hydro}
\rho(|x/\tau|\leq1)=\int_{k_F^-}^{k_F^+} \frac{\dd k}{2\pi}=\frac{k_F^+(x,\tau)-k_F^-(x,\tau)}{2\pi},
\ee
which reduces to Eq.~\eqref{dens-intro} in the case of an infinite-wall, see \cite{Antal1999,Antal2008}.

\begin{figure}[t!]
\centering
\includegraphics[width=\textwidth]{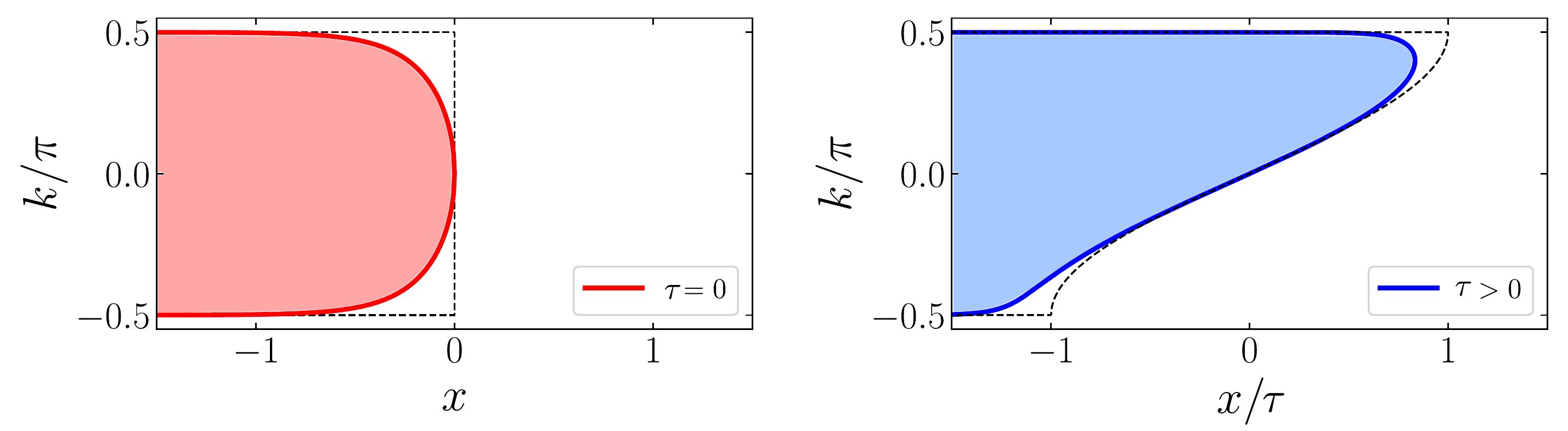}
\caption{\label{fig:W-t-HW}{(Left panel)} Wigner function for the ground state of the confined gas in Eq.~\eqref{xx-confined-continuum}  at $\tau=0$ \eqref{W-0} and {(right panel)} for $\tau>0$ \eqref{W-t}.  The colored regions show the Wigner function for the potential \eqref{exp-potential} while the dashed lines show the contour of the Wigner function for an infinite-wall confinement \eqref{potential}.}
\end{figure}

Such semi-classical description of the gas at hydrodynamic scales can be equivalently described in terms of the occupation number $W(x,k)$ of each Fourier mode $\hat\eta^\dagger_{k,x}$ inside the fluid cell $[x,x+\Delta x]$, which is nothing but the Wigner function of the non-interacting Fermi gas \cite{Wigner1997}, see also \cite{Hinarejos2012}. It is defined in terms of the fermionic creation and annihilation operators as
\be
W(x,k)=\int_0^{\Delta x} \frac{\dd y}{\delta \ \Delta x}\int_{-\infty}^\infty \dd q \  e^{\I k q} \; \braket{ \Cc^\dagger_{\nicefrac{(x+y+q)}{2}}\Cc_{\nicefrac{(x+y-q)}{2}}}
\ee
and it carries the physical interpretation of the semi-classical probability of finding a particle in a coarse-grained position $(x,k)$ of the phase space. In terms of the Wigner function, the ground state of the confined gas at $\tau<0$ reads
\be\label{W-0}
W(x,k)=\begin{cases} 1 \qquad\text{if $|k|\leq  \arccos V(x)$}, \\[4pt] 0 \qquad\text{otherwise}.\end{cases}
\ee
At $\tau >0$, the time-evolution of the Wigner function is given by the Moyal equation \cite{Fagotti2017,Fagotti2020,Moyal1949} and, at lowest order in the $\de_x$ and $\de_k$ derivatives, it satisfies the transport equation\cite{Ruggiero2019}
\begin{equation}
\partial_\tau W(x,k,\tau)+\sin k\ \partial_xW(x,k,\tau)=0
\end{equation}
with solution
\be\label{W-t}
W(x,k,\tau)= W(x-\tau\sin k,k,0),
\ee
which essentially implements the constraint of Eq.~\eqref{eq-of-motion}. The Wigner function approach allows for a graphical interpretation of the gas expansion in phase-space, as depicted in Fig.~\ref{fig:W-t-HW}. 
Notice that the problem under analysis is characterized by the presence of only two Fermi points $k_F^\pm(x,\tau)$ at each time $\tau$ and around each spatial position $x$. 
This can be seen from Eq.~\eqref{self-const} but also from Fig.~\ref{fig:W-t-HW}, as any vertical line will have either zero or two intersections with the contour of the Wigner function.  Problems for which there are more than two Fermi points have also been considered, see \cite{Ruggiero2020} and references therein. 

%%%%%%%%%%%%%%%%%%%%%%%%%%
%%%%%%%%%%%%%%%%%%%%%%%%%%

\subsection{Quantum fluctuating hydrodynamics}\label{sec:quantum-fluct-hydro}
We now consider quantum fluctuations around the classical hydrodynamic description presented in Sec.~\ref{sec:hydro-classical}. For a given time $\tau$ and position $x$, it is possible to describe the quantum fluctuations around a classical configuration by introducing the density fluctuating-field $\hat\phi$ \cite{Brun2017,Brun2018,Ruggiero2019}
\be
\delta\hat\rho(x,\tau)=\frac{1}{2\pi}\ \de_x \hat\phi(x,\tau)
\ee
in such a way that the semi-classical density profile in Eq.~\eqref{dens-hydro} accounts for particle-hole pairs formation around the Fermi points. In fact, such processes dominate the low-energy quantum physics of the model at the large scales \cite{Cazalilla2004,Giamarchi2007}. Moreover, by applying standard quantum hydrodynamics techniques \cite{Giamarchi2007}, it is possible to identify the time-dependent fermionic creation and annihilation fields $\Cc^\dagger_x(\tau)$, $\Cc_x(\tau)$ with a sum of local operators in the low-energy theory which share the same symmetry of the initial model. Following this program and  retaining only the leading order terms in the sum (i.e., those operators in the low-energy theory with smallest scaling dimension), one obtains, up to a non-universal amplitude, the following expansion \cite{Allegra2016,Dubail2017,Ruggiero2019,Scopa2020}
\be\begin{matrix}
\Cc^\dagger_x(\tau) \propto  e^{\I \varphi(x,\tau)} \ \textbf{:} \exp\left(\frac{\I}{2}\left[ \hat\phi_+(x,\tau) -\hat\phi_-(x,\tau)\right]\right) \textbf{:} + \ \text{less relevant operators}\\[10pt]
\Cc_x(\tau) \propto  e^{-\I \varphi(x,\tau)} \ \textbf{:} \exp\left(\frac{\I}{2}\left[ \hat\phi_-(x,\tau) -\hat\phi_+(x,\tau)\right]\right) \textbf{:} + \ \text{less relevant operators}
\end{matrix}
\ee
 where $\text{\bf :}\ . \ \text{\bf :}$ denotes the normal ordering of operators and $\varphi=\frac{1}{2}(\varphi_+-\varphi_-)$ is a semi-classical phase. 
 {The latter} is obtained by integrating the differential phase
\be
\dd \varphi_{\pm}(x,\tau)= k_F^\pm(x,\tau) \dd x - \varepsilon(k_F^\pm(x,\tau),x) \dd t,
\ee
which is the phase carried at small distances $\dd x$ and at small times $\dd t$ by the single-particle wave-function after the creation of a particle at position $(x,k_F^\pm)$ in phase space. The fields $\hat\phi_\pm(x,\tau)$ are instead the chiral components of $\hat\phi=\hat\phi_-+\hat\phi_+$ and carry the physical interpretation of left- and right- moving parts of the density fluctuating-field.
The low-energy effective Hamiltonian that governs the dynamics of the quantum fluctuations is that of a non-homogeneous Luttinger liquid (see e.g. \cite{Allegra2016,Dubail2017,Brun2017,Brun2018,Scopa2020,Bastianello2020})
\begin{equation}\label{LL}
	\Ha_{\rm LL}=  \frac{1}{2\pi} \int_{-L}^\tau \dd x \  \left[\sin k_F^+(x,\tau) \; \left(\de_x \hat\phi_{a_+}(x,\tau) \right)^2+ \sin k_F^-(x,\tau) \; \left(\de_x \hat\phi_{a_-}(x,\tau) \right)^2\right],
 \end{equation}
whose dependence on the spatially dependent Fermi velocity can be removed with a simple change of coordinate $x\to \tilde{x}$, see Sec.~\ref{sec:param}. Here, $a_\pm\equiv a(k_F^\pm(x,\tau))$ and $a(k)=\mp$ if ${\rm sign}(k)\lessgtr 0$.\\

 Equivalently, introducing a parametrization of  the initial contour of local Fermi-points \eqref{kF-initial} (or {\it Fermi contour}) with the coordinate $\theta$ along the curve 
\be\label{Fermi-contour}
\Gamma=\left\{ \big(x(\theta),k(\theta)\big)\, : \, \left|k(\theta)\right|=\arccos V\big(x(\theta)\big)\right\} 
\ee
one can write the Luttinger liquid Hamiltonian as
\be\label{LL-contour}
\Ha_{\rm LL}[\Gamma]=\int_\Gamma \frac{\dd\theta}{2\pi} \ {\cal J}(\theta) \ \sin k(\theta) \, \left(\de_\theta \ \hat\phi_a(\theta) \right)^2,
\ee
where ${\cal J}(\theta)$ is a Jacobian factor and $a=a(k(\theta))$.
In particular, in our hydrodynamic description of the problem, quantum fluctuations in the initial state are given by the ground state of $\Ha_{\rm LL}[\Gamma]$. As time flows, the Fermi contour $\Gamma$ is modified according to \eqref{W-t} and the quantum fluctuations are transported along the curve, spreading the entanglement {through the system}.

\section{Entanglement entropy evolution during the gas expansion}\label{sec:entanglement}
We now investigate the evolution of the entanglement entropy for the setup of Sec.~\ref{sec:model}. Notice that, although we deal with a non-interacting system, the derivation of exact results for the entanglement is currently out-of-reach with a lattice formulation of the setup under analysis. 
Conversely, in the hydrodynamic approximation, such non-equilibrium and non-homogeneous problems \cite{Dubail2017,Collura2020} can be easily handled with tools stemming from CFT.
%In its lattice formulation, the derivation of exact results for the entanglement entropy evolution is, in general, complicated and limited to simple settings but, in hydrodynamic approximation, it can be handled with tools stemming from CFT, allowing also for the asymptotic investigation of non-equilibrium inhomogeneous problems \cite{Dubail2017,Collura2020}.

In particular, in the continuum limit, the $\alpha$-R\'enyi entropy for integer $\alpha$ can be related to the following expectation value of the twist field $\hat{\cal T}_\alpha$ \cite{Calabrese2004,Cardy2008,Calabrese2009}  
\be\label{EE}
\tilde{S}_\alpha(x,\tau)= \frac{1}{1-\alpha} \log\braket{\hat{\cal T}_\alpha(x,\tau)}.
\ee
Moreover, in our model, the twist field can be written as product of the two chiral twist-fields $\{ \hat\Phi^+_\alpha, \hat{\Phi}^-_\alpha\} $ and, under conformal mappings, such chiral fields behave as primary CFT operators with dimension
\be
h_\alpha=\frac{c}{24}\left(\alpha -\frac{1}{\alpha}\right)
\ee
where $c$ is the central charge of the underlying CFT,  in particular $c=1$ for free Fermi gases. 

\begin{figure}[t!]
\centering
\includegraphics[width=\textwidth]{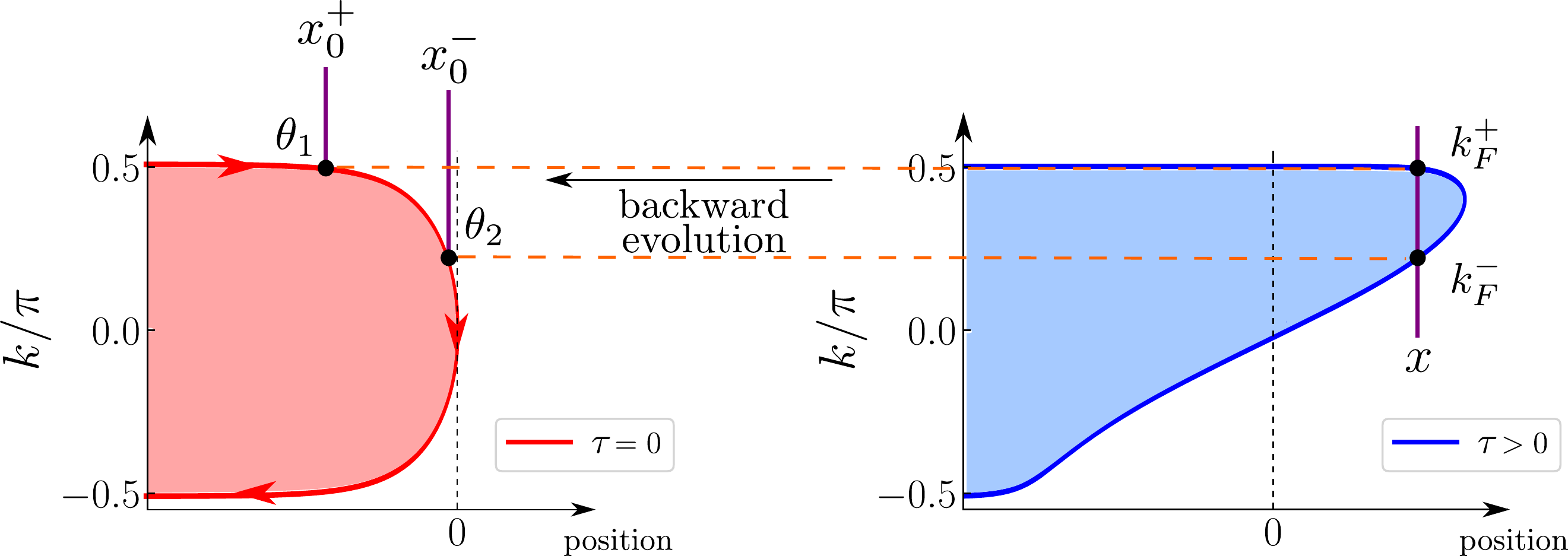}
\caption{\label{fig:backwards-evo} Position of the Fermi points $k_F^\pm(x,\tau)$ in terms of the coordinate $\theta$ along the initial Fermi contour. (Right) At time $\tau>0$ the coordinates of the Fermi points are $(x,k_F^\pm)$ in the $k$-$x$ space while (left) at $\tau=0$ they can be traced back to their initial positions $(x_0^\pm,k_F^\pm)$ and expressed in terms of the coordinate $\theta_{1,2}$ along the curve.}
\end{figure}

It follows that  $\tilde{S}_\alpha(x,\tau)$ can be written as the two-point correlation function of the chiral fields $\hat\Phi^+_\alpha$, $\hat{\Phi}^-_\alpha$ of the CFT which live along the Fermi contour at time $\tau$.  
The Fermi points $k_F^\pm(x,\tau)$ at time $\tau$ are traced back to the initial Fermi contour where they are found at positions $x^\pm_0$. 
Therefore, if $\theta$ denotes a parametrization of the  contour $\Gamma$ \eqref{Fermi-contour}, then the computation of the $\alpha$-R\'enyi entropy reduces to 
\be\label{entanglement-hydro} 
\tilde{S}_\alpha(x,\tau)=\frac{1}{1-\alpha} \log\left(\left|\frac{\dd\theta}{\dd x}\right|^{h_\alpha}_{\theta=\theta_1}\left|\frac{\dd\theta}{\dd x}\right|^{h_\alpha}_{\theta=\theta_2} \braket{\hat\Phi^+_\alpha(\theta_1) \hat{\Phi}^-_\alpha(\theta_2)}\right),
\ee 
where $\theta_{1,2}$ are the coordinates of $k_F^\pm(x,\tau)$ along the initial Fermi contour, see Fig.~\ref{fig:backwards-evo}. 
%%%%%%%%%%%%%%%%%%%%%%%%%%
%%%%%%%%%%%%%%%%%%%%%%%%%%
%

We emphasize that Eq.~\eqref{EE} provides solely the universal large-scale contribution to the $\alpha$-R\'enyi entropies and it has to be complemented with a non-universal cutoff so that 
\be\label{EE-full}
S_\alpha(x,\tau)= \frac{1}{1-\alpha} \log\left[ \epsilon^{2h_\alpha} \braket{\hat{\cal T}_\alpha(x,\tau)}\right]=\tilde{S}_\alpha(x,\tau) + \frac{2h_\alpha}{1-\alpha}\log \epsilon(x,\tau),
\ee
which {encodes also the} ultraviolet divergences of $\tilde{S}_\alpha$ \cite{Calabrese2004,Calabrese2010}. 
In our hydrodynamic approach, 
the only relevant microscopic scale is the inverse local density $\rho^{-1}$, see Eq.~\eqref{dens-hydro}, { which in the von Neumann entropy enters} as \cite{Jin2004,Calabrese2010}
\be\label{cutoff}
\epsilon(x,\tau)=\frac{C}{\sin\pi\rho(x,\tau)} 
\ee
where $C$ is a non-universal constant related to $\Upsilon$ appearing in the {anticipated final results for the} 
entanglement entropy \eqref{solution} as $\Upsilon=-\frac{1}{6}\log C$. %Its structure and its derivation can be found in \cite{Jin2004}.
%%%%%%%%%%%%%%%%%%%%%%%%%%%%%%%
%%%%%%%%%%%%%%%%%%%%%%%%%%
\subsection{Parametrization of the initial Fermi contour}\label{sec:param}

Since the modes propagate at different velocities, we first introduce the stretched coordinate \cite{Allegra2016,Dubail2017,Brun2017,Brun2018,Ruggiero2019,Bastianello2020,Ruggiero2020,Scopa2020}
\be\label{isothermal}
\begin{split}
\tilde{x}(x_0)&=\int_{-L}^{x_0} \frac{ \dd x'}{\sin k_F(x')} \\
&=\int_{-L}^{x_0} \frac{ \dd x'}{\sqrt{1-V(x')^2}}  \\
&= -\frac{1}{\beta}\tanh^{-1}\sqrt{1- e^{2\beta x'}} \ \Bigg\vert_{-L}^{x_0}
\end{split}
\ee
which measures the time needed by an excitation emitted at position $-L$ to reach the position $x_0$ with spatially-dependent propagation velocity $\sin k_F(x)$.

\begin{figure}[t!]
\centering
\includegraphics[width=0.6\textwidth]{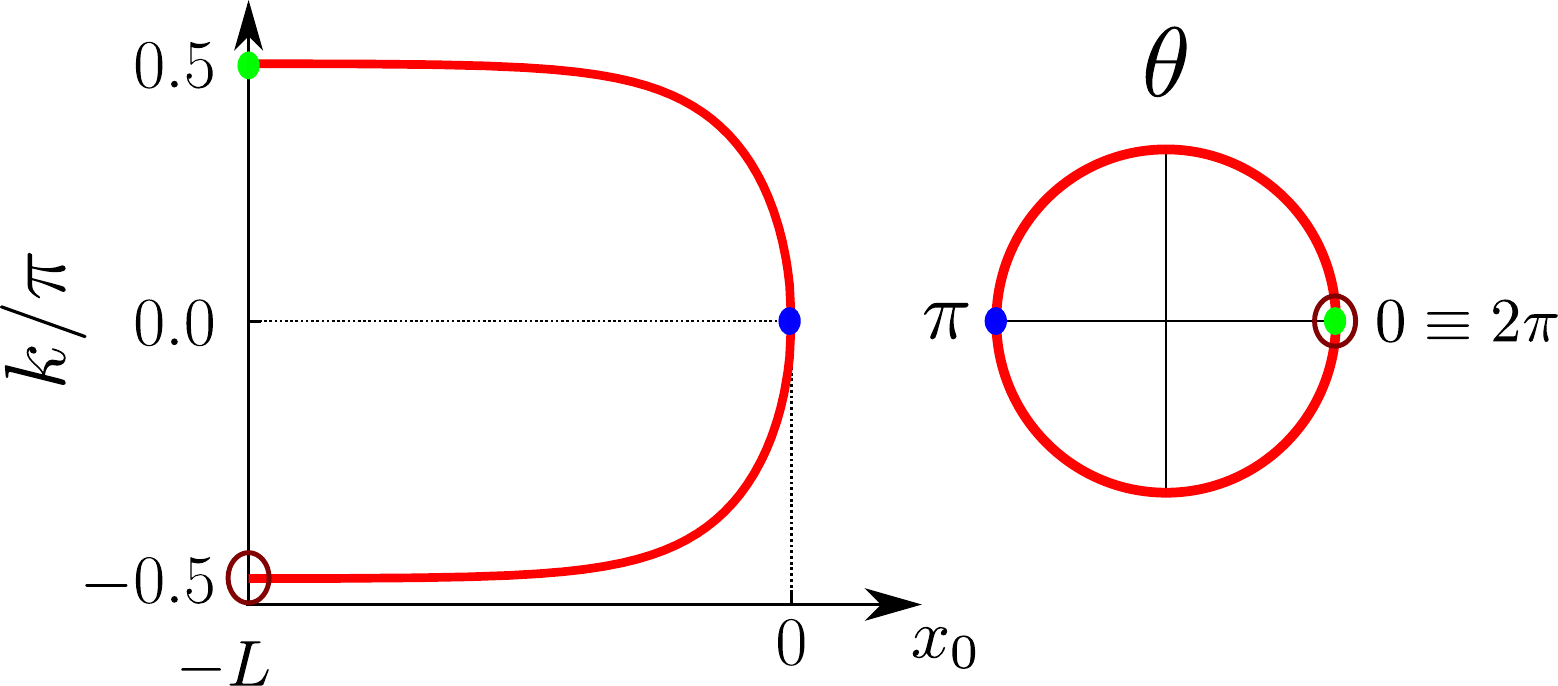}
\caption{\label{fig:isothermal} Illustration of the parametrization of the initial Fermi contour. This procedure may be viewed as a stretching (see Eq.~\eqref{isothermal}) and a subsequent folding (see Eq.~\eqref{param}) of the initial Fermi contour along the unit circle. In this way, the reflection of the modes at the boundary $x=-L$ are properly accounted in the CFT by identifying the angles $\theta=\theta \ {\rm mod}\ 2\pi$.}
\end{figure}

Equation~\eqref{isothermal} provides an unambiguous parametrization of the upper Fermi contour and can be easily extended to the lower part with symmetry arguments. In particular, we define the angular variable
\be\label{param}
\theta(x_0)= \begin{cases}
\frac{\pi}{\cal N} \tilde{x}(x_0), \qquad &\text{if $\arcsin\frac{x-x_0}{\tau}\geq 0$}\\[4pt]
\pi + \frac{\pi}{\cal N} \tilde{x}(x_0) \qquad &\text{otherwise}.
\end{cases}
\ee
The normalization  ${\cal N}$ of the stretched coordinate is given by
\be
{\cal N}= \int_{-L}^0 \frac{ \dd x'}{\sin k_F(x')}=\frac{1}{\beta}\tanh^{-1}\sqrt{1-e^{-2\beta L}} \  .
\ee
The second line in Eq.~\eqref{param} may be viewed as the folding of the stretched coordinate along the unit circle so that the finite-size of the chain $L$ is properly accounted in the CFT along the Fermi contour, see Fig.~\ref{fig:isothermal}. \\
In terms of the coordinate $\theta$, the low-energy effective Hamiltonian \eqref{LL-contour} takes the Jacobian factor ${\cal J}=\frac{\pi}{\cal N} [\sin k_F]^{-1}$, which cancels out the dependence of $\Ha_{LL}[\Gamma]$ on the non-homogeneous Fermi velocity.\\

%%%%%%%%%%%%%%%%%%%%%%%%%%
%%%%%%%%%%%%%%%%%%%%%%%%%%

%

Equipped with the parametrization \eqref{param} of the initial Fermi contour \eqref{Fermi-contour}, we now evaluate the entanglement entropy. At given positions in space $x$ and in time $\tau$, the large-scale contribution to the entanglement profile is obtained from Eq.~\eqref{entanglement-hydro} with the two-point correlation function along the unit circle
\be
\braket{\hat\Phi^+_\alpha(\theta_1)\hat{{\Phi}}^-_\alpha(\theta_2)}=\left(\sin\frac{\theta_1-\theta_2}{2}\right)^{-2h_\alpha},
\ee
$\theta_{1,2}=\theta(x_0^\pm)$, and the Weyl factors
\be
\begin{split}
\left|\frac{\dd \theta(x_0)}{\dd x}\right|_{x_0=x^\pm_{0}(x,\tau)} &= \frac{\pi}{{\cal N} }\tilde{x}^\prime(x_0^\pm(x,\tau))  \left|\frac{\dd x^\pm_0(x,\tau)}{\dd x}\right|\\
&=\frac{\pi}{{\cal N} }\frac{1}{\sqrt{1-V( x_0^\pm(x,\tau))^2}}  \left|\frac{\dd x_0^\pm(x,t)}{\dd x}\right|
\end{split}
\ee
where the final Jacobian is obtained from Eq.~\eqref{self-const} as 
\be
\frac{\dd x_0}{\dd x}=\frac{x-x_0}{x-x_0-\tau^2 V^\prime (x_0)V(x_0)}= \frac{x-x_0}{x-x_0-\beta \ \tau^2  \ e^{2\beta x_0}},
\ee
so that overall we have 
\begin{equation}
\begin{split}
\left|\frac{\dd \theta(x_0)}{\dd x}\right|_{x_0=x^\pm_{0}(x,\tau)} &=\frac{\pi}{{\cal N} } \left|\frac{x-x^\pm_{0}(x,\tau)}{t}-\beta \tau\left(1- \frac{(x-x^\pm_{0}(x,\tau))^2}{\tau^2}\right)\right|^{-1}.
\end{split}
\end{equation}
The cutoff $\epsilon(x,\tau)$ is {read off} from Eq.~\eqref{cutoff} as function of the particle density profile $\rho(x,\tau)$, which can be expressed in terms of 
$x_0^\pm(x,\tau)$ by {plugging} Eq.~\eqref{kF-def} into \eqref{dens-hydro}.
Combining Eqs.~\eqref{entanglement-hydro}, \eqref{cutoff} with Eq.~\eqref{EE-full} and taking the limit $\alpha\to 1$, we finally obtain
\be\label{expectation}
S_1(x,\tau)=\frac{1}{6}\log\left(\left|\frac{\dd\theta}{\dd x}\right|_{\theta=\theta_1}^{-1/2} \left|\frac{\dd\theta}{\dd x}\right|_{\theta=\theta_2}^{-1/2} 
\frac{1}{C}\left|\sin\frac{k_F^+-k_F^-}{2} \right|\left|\sin \frac{\theta_1-\theta_2}{2}\right|\right) 
\ee
that can be written for $\tau >0$ in terms of the ratios 
\be
\zeta_{1,2}= \frac{x-x_0^\pm(x,\tau)}{\tau}
\ee
and after simple algebra, as
\be\label{S-finite-beta}\begin{split}
%%%%%%%%%%%%%%%%%%%%%%%%%%%555
&S_1(x,\tau>0)=\frac{1}{6}\log\Bigg[\frac{\cal N}{2\pi}\sqrt{\left|\left(\zeta_1-\beta \tau (1-\zeta_1^2)\right)\left(\zeta_2-\beta \tau (1-\zeta_2^2)\right)\right|}
 \\[4pt]
&\Bigg|{\rm sign}(\zeta_1)\sqrt{\left(1+\sqrt{1-\zeta_2^2}\right)\left(1-\sqrt{1-\zeta_1^2}\right)}
-{\rm sign}(\zeta_2)\sqrt{\left(1+\sqrt{1-\zeta_1^2}\right)\left(1-\sqrt{1-\zeta_2^2}\right)}\Bigg|\\[4pt]
& \times\Bigg| \sin\left(\frac{\pi}{2\beta{\cal N}}\left( \tanh^{-1}|\zeta_2| -\tanh^{-1}|\zeta_1|\right) +\frac{\pi}{2}({\rm sign}(\zeta_1) -{\rm sign}(\zeta_2)) \right) \Bigg|\Bigg] +\Upsilon,
\end{split}\ee
where we {factorized} the non-universal constant $\Upsilon\equiv -\frac{1}{6}\log C\simeq 0.49502$ \cite{Jin2004}. Eq.~\eqref{S-finite-beta} gives the entanglement entropy evolution of a free Fermi gas initially {trapped in an} exponential potential \eqref{exp-potential} 
up to subleading  corrections {in time} (which {hereafter} we systematically drop). 
{The values of $\zeta_{1,2}$ are provided by} the numerical solutions for $x_0^\pm(x,\tau)$ in Eq.~\eqref{self-const}.

At $\tau=0$, the formula for the entanglement entropy simplifies since $k_F^\pm=\pm \arccos V(x)$ and $\theta_1=-\theta_2$, from which we arrive to the result
\be\label{S-finite-beta0}
S_1(x,0)=\frac{1}{6}\log\Bigg[{\frac{\cal N}{\pi}}\sqrt{1-V(x)^2} \;\Bigg| \sin\left(\frac{\pi}{\beta{\cal N}} \tanh^{-1}\sqrt{1-V(x)^2}\right) \Bigg|\Bigg] +\Upsilon.
\ee

\begin{figure}[t!]
\centering
\includegraphics[width=\textwidth]{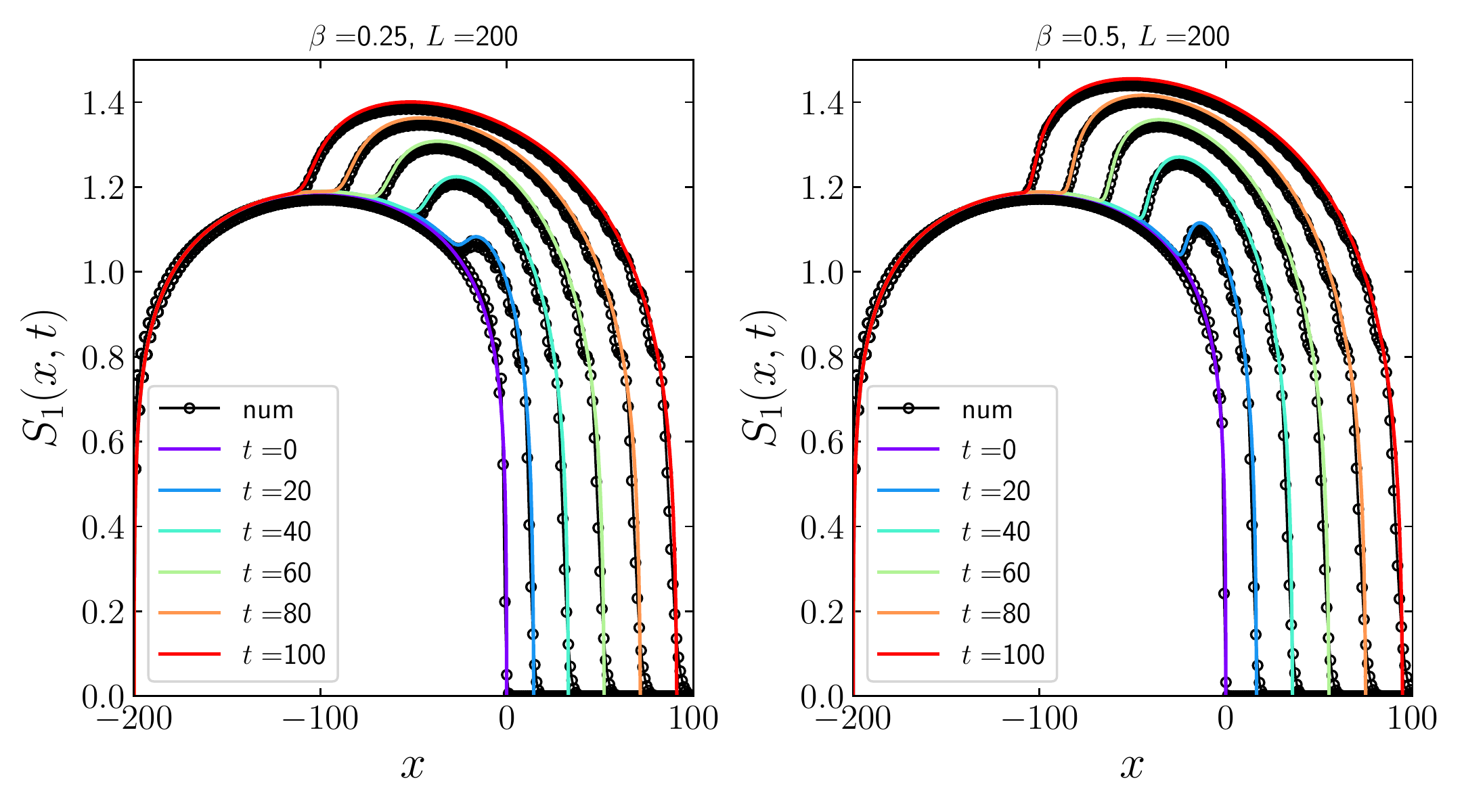}
\caption{\label{fig:S-beta} Evolution of the entanglement profiles in Eq.~\eqref{S-finite-beta} for the trap-release protocol of the gas in Eq.~\eqref{xx-confined-continuum} with potential \eqref{exp-potential} (colored curves) compared with exact numerics (circles) for the lattice Hamiltonian \eqref{xx-lattice-exp}. The plots show the entanglement profiles at different times $t=\tau/\delta$ (measured in units of $\delta\equiv 1$) and for different values of $\beta=0.25,0.5$. The agreement of the curves with numerics is excellent.}
\end{figure}

 %Note that the entanglement entropy can be entirely expressed in terms of the ratios $\zeta_{1,2}$, which describe the rescaled distance traveled by the quasi-particles giving rise to the correlations.
Although Eqs.~\eqref{S-finite-beta} and\eqref{S-finite-beta0} give the entanglement profiles for a trap-release protocol that {generalize} the setup of Sec.~\ref{sec:model} 
{with the presence of a} finite value of $\beta$, {it is still an interesting physical protocol}. 
Therefore, we proceed to test of Eq.~\eqref{S-finite-beta} and \eqref{S-finite-beta0} with exact numerical computations for the lattice Hamiltonian \eqref{free-gas} 
of size $2L$ with exponential potential for $t<0$:
\be\label{xx-lattice-exp}
\Ha= -\frac{1}{2}\sum_{j=-L}^{L-2} \left(\Cc_j^\dagger \Cc_{j+1} + \Cc_{j+1}^\dagger \Cc_j \right)  \, + \, \sum_{j=-L}^{L-1} \exp(\beta j \delta)\ \Cc_j^\dagger \Cc_{j} .
\ee 
In particular, we {first compute the two-point correlation matrix in the ground state of the initial Hamiltonian \eqref{xx-lattice-exp} and we subsequently exactly evolve it with} 
 post-quench Hamiltonian (where $V=0$). For free Fermi gases, the entanglement entropy {is obtained with 
%\be
%S_1(j,t)=\sum_{n=1}^{L-j}\left[ \xi_n(t) \log \xi_n(t) - (1-\xi_n(t)) \log(1-\xi_n(t))\right]
%\ee
%where $\xi_n$ are the eigenvalues of the two-point correlation matrix restricted to the subsystem $A=[j,L]$, see 
the techniques of \ref{app:numerics}}. 
The results are reported in Fig.~\ref{fig:S-beta} for different times and for different values of $\beta=0.25,0.5$. 
The agreement of the {analytic prediction} with the numerics is extremely good.
%%%%%%%%%%%%%%%%%%%%%%%%%%
%%%%%%%%%%%%%%%%%%%%%%%%%%

%

We {stress} that the result in Eq.~\eqref{S-finite-beta} is independent {from the employed} parametrization of the initial Fermi contour. {As a pedagogical example},  we report in \ref{app:param-Jerome} the computation of the entanglement entropy profiles for a different (but equivalent) parametrization 
of the initial Fermi contour which leads to the same final result.

\subsection{Entanglement entropy profiles in the limit $\beta\to \infty$ \label{sec:solution}}
At this point, we {are finally ready to take}  the limit of large $\beta$ of the exponential trap-release protocol studied {above}. 
As discussed in Sec.~\ref{sec:hydro-limit}, such auxiliary problem asymptotically reproduces the free expansion setup of Sec.~\ref{sec:model} 
{which is the main goal of this paper}. 

We first notice that Eq.~\eqref{self-const} can be solved analytically at large $\beta$. In particular, {as long as} $x/\tau\leq-1$ we find the following roots
\be
x_0^-= x+\tau;\qquad  x_0^+=x-\tau
\ee
up to exponentially small corrections, whereas, when $|x|/\tau< 1$, one finds the two solutions
\be
x_0^-=\frac{1}{2\beta}\log(1-\frac{x^2}{\tau^2}), \qquad x_0^+=x-\tau.
\ee
{In the latter regime}, the corrections to the solution $x_0^-$ decay algebraically in $1/\beta$.

Introducing the stretched coordinate of Eq.~\eqref{isothermal}, one {easily} sees that the normalization constant {is} ${\cal N}=L$ up to 
$\sim{\cal O}(1/\beta)$ corrections, {because} the region where $|k_F|\neq \pi/2$ has a zero measure for large $\beta$. 
It follows that the angular coordinate~\eqref{param} (in the upper branch of the initial Fermi contour) can be written as
\be
\theta(x_0)=\begin{cases}
\pi+ \frac{\pi}{L} (x+ \tau), \qquad &\text{if $x/\tau\leq-1$ and $x_0=x_0^-$};\\[4pt]
\pi- \frac{\pi}{L\beta}\tanh^{-1}(x/\tau), \qquad &\text{if $|x|/\tau< 1$ and $x_0=x_0^-$}; \\[4pt]
\pi+ \frac{\pi}{L} (x- \tau), \qquad &\text{if $x_0=x_0^+$},
\end{cases}
\ee
from which one can derive the Weyl factor
\be\label{weyl-beta-infty}
\left|\frac{\dd\theta(x_0)}{\dd x}\right|=\begin{cases} \frac{\pi}{\beta L} \left[\tau(1-\frac{x^2}{\tau^2})\right]^{-1}, \qquad &\text{if $|x|/\tau< 1$ and $x_0=x_0^-$;}\\[4pt]
\pi/L,\qquad &\text{otherwise.}
\end{cases}
\ee
The density profile is easily derived from Eqs.~\eqref{kF-def}-\eqref{dens-hydro} as
\be\label{rhoHW}
\rho(x,\tau)=\begin{cases}  \arccos(x/\tau)/(2\pi), \qquad &\text{if $|x|/\tau< 1$};\\[4pt]
1/2,\qquad &\text{if $x/\tau\leq-1$};\\[4pt]
0, \qquad &\text{otherwise}
\end{cases}
\ee
and it reproduces the expected value in Eq.~\eqref{dens-intro}.

 At this point, {starting} from Eq.~\eqref{expectation} and {after} simple algebra, one finds the entanglement entropy profile. 
 In particular, for $x\leq-\tau$, {we get}
\be\label{S-out}
S_1(x\leq-\tau)=\frac{1}{6}\log\left({\frac{L}{\pi}}\left|\sin \frac{\pi x}{L}\right|\right) +\Upsilon
\ee
whereas, from Eq.~\eqref{S-finite-beta}, {we obtain} for $|x|< \tau$
\be\label{full}\begin{split}
S_1(|x|<\tau)=& \frac{1}{6}\log\Bigg(\frac{L}{2\pi}\sqrt{\left|\frac{x}{\tau} -\beta \tau(1-\frac{x^2}{\tau^2})\right|} \ \left|\sqrt{1+\sqrt{1-\frac{x^2}{\tau^2}}}-{\rm sign}(x)\sqrt{1-\sqrt{1-\frac{x^2}{\tau^2}}}\right|  \\
&\times \left|\sin\Big(\frac{\pi}{2L}(x-\tau)\Big) \right| \Bigg) +\Upsilon.
\end{split}\ee
%After noticing that Eq.~\eqref{full} matches with \eqref{S-out} at position $x=-\tau$ (i.e., the entanglement profile is a continuous function at the light-cone), one can further approximate Eq.~\eqref{full} to the leading order {in $\beta$} as
%\be\begin{split}\label{leading}
%S_1(|x|<\tau)=& \frac{1}{6}\log\Bigg(\frac{L}{2\pi}\sqrt{\beta \tau(1-\frac{x^2}{\tau^2})} \ \left|\sqrt{1+\sqrt{1-\frac{x^2}{\tau^2}}}-{\rm sign}(x)\sqrt{1-\sqrt{1-\frac{x^2}{\tau^2}}}\right|  \\
%&\times \left|\sin\Big(\frac{\pi}{2L}(x-\tau)\Big) \right| \Bigg)+\Upsilon.
%\end{split}\ee
Finally, recalling that asymptotically $\beta \sim 1/\delta$ (see Eq.~\eqref{beta-limit}), one obtains the result {anticipated} in Eq.~\eqref{solution}, 
where the time $t$ is measured in units of $\delta$ and $L\to \infty$. 

The result in Eq.~\eqref{solution} has been tested against exact numerics made for the lattice Hamiltonian \eqref{free-gas} of size $2L$ with a infinite-wall confinement \eqref{potential}. In Fig.~\ref{fig:summary}{\bf(b)} we see that the solution \eqref{solution} correctly reproduces the evolution of the entanglement profiles for the setup of Sec.~\ref{sec:model}, even for modest value of the system size $L$, set to $200$ in our numerical calculations.

%%%%%%%%%%%%%%%%%%%%%%%%%%%%
\subsection{Asymptotic growth of the entanglement}
\begin{figure}[t!]
\centering
\includegraphics[width=\textwidth]{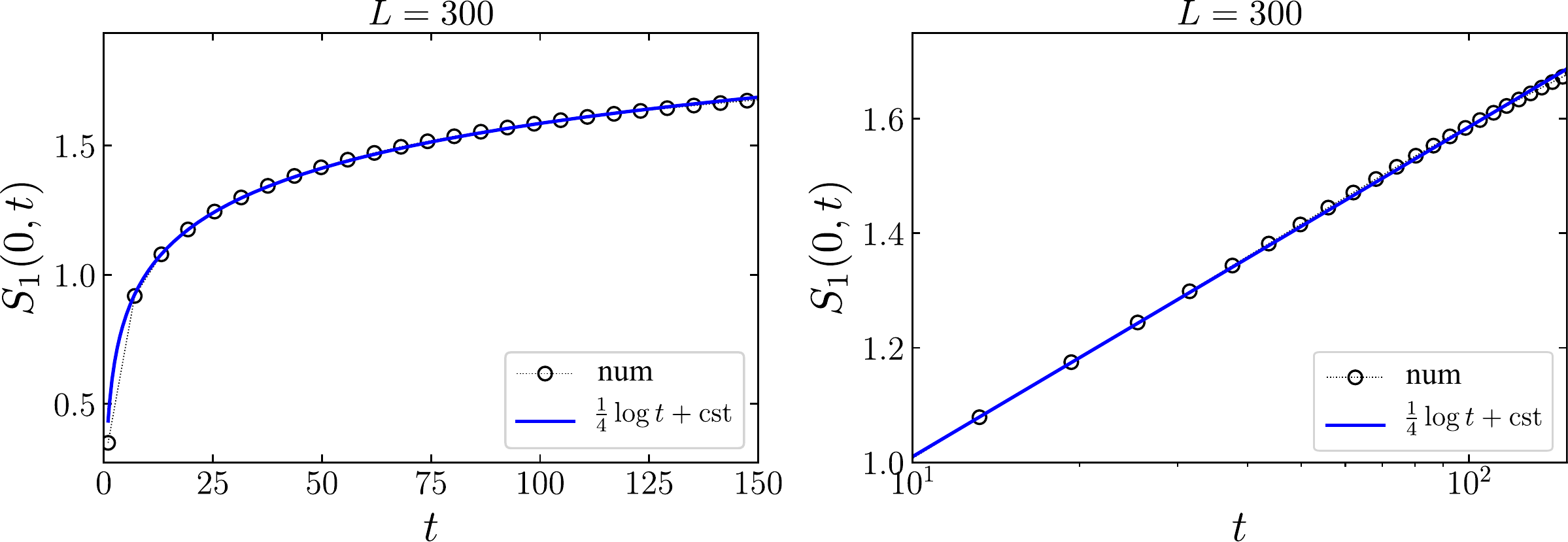}
\caption{\label{fig:S-asy} Time evolution of the {half-system entanglement entropy (i.e. $j=0$)} for large times $t\ll L$. 
{(Left)} The asymptotic prediction~\eqref{asy-prediction} is tested against exact numerics for the Hamiltonian \eqref{free-gas} with potential \eqref{potential} and size $2L$. 
{(Right)} Same {plot} in logarithmic scale. }
\end{figure}

It is {worth to briefly discuss} the time evolution of the entanglement entropy {for} $j=0$. 
{Starting} from Eq.~\eqref{full} and expanding the sine for $L\gg j,t$, we obtain at leading order and up to an additive constant
\be
S_1(|j|<t)\sim \frac{1}{6}\log\left(\sqrt{{\left|\frac{j}{t}-t(1-\frac{j^2}{t^2})\right|}} |j-t| \right),
\ee
which reproduces  Eq.~\eqref{asy-prediction} upon setting $j=0$. 
{Notice in particular how the argument of the logarithm conspires to give $t^{3/2}$ which is the origin of the unusual $\frac14 \log t$ growth.   
Exact numerical calculations for} the lattice model, presented in Fig.~\ref{fig:S-asy}, {confirm} that the {half-system} entanglement entropy {behaves as} $\frac{1}{4} \log t$, 
as expected {also} from previous results in literature \cite{Alba2014,Gruber2019}.

%%%%%%%%%%%%%%%%%%%%%%%%%%
%%%%%%%%%%%%%%%%%%%%%%%%%%

Concluding, {we recall a simple argument pointed out in Ref.~\cite{Gruber2019} for the asymptotic entanglement growth \eqref{asy-prediction}}.
The entanglement entropy during the gas expansion {can be thought as} the sum of two contributions. 
The first is simply the propagation of particles towards the right vacuum and it gives the standard $\sim\frac{1}{6}\log t$ growth of the entanglement entropy, 
{observed in the standard domain-wall quench} \cite{Dubail2017}. 
The second (which is the main peculiarity of this setup) comes from the spreading of the initial correlations of the gas 
and {its contribution can be estimated} as $\sim \frac{1}{12}\log t$ \cite{Gruber2019}. 
Thus, summing up the two, one eventually finds that the entanglement asymptotically grows as $\sim\frac{1}{4}\log t$.

\section{Summary and conclusion\label{sec:conclusion}}
We investigated the entanglement entropy evolution after a quantum quench of a one-dimensional gas of hard-core particles at zero-temperature and at half-filling. The gas is initially confined inside the interval $[-L,0]$ by two {infinite walls}. At $t=0$ the wall at position $j=0$ is suddenly removed and the gas is let free to expand, see Fig.~\ref{fig:summary}({\bf a}). The non-equilibrium dynamics of the gas is characterized by a non-homogeneous density profile which can be understood, in simple terms, with a semi-classical hydrodynamic approach, see Sec.~\ref{sec:hydro-classical}. The study of the entanglement is instead non-trivial and it requires the use of quantum fluctuating hydrodynamics and of conformal field theory, see Sec.~\ref{sec:quantum-fluct-hydro} and Sec.~\ref{sec:entanglement}. 
This program eventually leads to an exact prediction for the entanglement entropy evolution \eqref{solution} which is found in perfect agreement with {exact numerics}.
{We stress that our hydrodynamics prediction applies almost as it is to a hard-core gas in the continuum (with minor modification to account for the filling), 
which can be simply studied adapting the overlap matrix technique of Refs. \cite{cmv-11,cmv-11b}.}

This work opens the doors to several further studies. For instance, it would be interesting to compute, with the same strategy, the entanglement entropy of a finite interval $[x_1,x_2]$. Indeed, such analysis will reduce to the computation of a four-point function of chiral twist fields, whose structure is known from CFT. Another example is to study a {\it quantum Joule expansion}, where at time $t=0$ the initial box potential is extended to size $[-L,L]$. In this context, it will be interesting to investigate the long-time dynamics of the entanglement, by considering the reflections of chiral modes at positions $j=\pm L$ \cite{Dubessy2020}. {Similar quenches can be studied in free fermionic Hamiltonians without particles conservation like the quantum {\rm\small XY} model \cite{Lieb1961}, see e.g. \cite{Mukherjee2007,Barmettler2010,Lancaster2010,Yoshinaga2021}.
}
Finally, this paper raises also some new questions. Perhaps, the most natural one is to ask what happens if the same setup is considered for an interacting quantum gas. This has been recently investigated for the case of a domain-wall melting \cite{Collura2020} but, in principle, it can be similarly done for this setup by using quantum generalized hydrodynamics \cite{Ruggiero2020,next-pub}.

%%%%%%%%%%%%%%%%%%%%%%%%%%%%%%%
\vspace{1cm}
{\bf Acknowledgments}.  PC, AK and SS acknowledge support from ERC under Consolidator grant number 771536 (NEMO). JD acknowledges support from {\it CNRS International Emerging Actions} under the grant QuDOD. SS acknowledges Paola Ruggiero and Sara Murciano for useful discussions. JD and SS acknowledge the international conference {\it Statistical Physics and Low Dimensional Systems} 2020 (Pont-\'a-Mousson) during which part of this work has been discussed.

\appendix
\section{Numerical methods}\label{app:numerics}
The entanglement entropy of a bipartition $A\cup B$ is defined as 
\be\label{ee-2}
S_A=-{\rm tr}\ \rrho_A   \log  \rrho_A
\ee
where $\rrho_A$ is the reduced density matrix of the subsystem $A$. For free Fermi models, the latter can be expressed in terms of an entanglement Hamiltonian $\hat{\cal H}$ \cite{p-12,Peschel1999a,Chung2001,Peschel2003,Peschel2004,Peschel2009}
\be
\rrho_A \propto \exp  \hat{\cal H}
\ee
which has spectrum
\be
e_j = \log \frac{ 1-\xi_j}{\xi_j}, \qquad {j=1\dots  |A|}.
\ee
Here, $\xi_j$ are the eigenvalues of the two-points correlation matrix $\mathbb{G}$ restricted to the subsystem $A$ {of length $|A|$}. 
{Hence,  the entanglement entropy~\eqref{ee-2} in terms} of either the eigenvalues of the entanglement Hamiltonian or the two-point correlation function is
\be\label{EE-latt}
S_A=\sum_{j=1}^{|A|}\left[ \log(1+e^{-e_j}) + \frac{e_j}{1+e^{e_j}}\right]=\sum_{j=1}^{|A|} \left[ \xi_j \log\xi_j - (1-\xi_j)\log(1-\xi_j)\right].
\ee 

{For our problem, the time-dependent correlation matrix is easily written down.  
We first write} the free Fermi Hamiltonian \eqref{free-gas} with $V_j=0$
\be\label{xx-matrix}
\Ha=\sum_{i,j} \Cc_j^\dagger \ \mathbb{H}_{i,j} \ \Cc_j, \qquad \mathbb{H}_{ij}=-\frac{1}{2}(\delta_{i+1,j} + \delta_{i,j+1}) 
\ee
and we require that the initial state $\ket{\Psi_0}$ is the ground state of the confined Hamiltonian \eqref{free-gas} 
\be\label{H0-latt}
\mathbb{H}^{(0)}_{ij}=-\frac{1}{2}(\delta_{i+1,j} + \delta_{i,j+1})+V_j\ \delta_{i,j}.
\ee
Next, we diagonalize the Hamiltonian matrices in \eqref{xx-matrix} and \eqref{H0-latt} as 
\be
\mathbb{H}=\bm{w}^\dagger  \ {\rm diag}(\bm\varepsilon) \ \bm{w}; \qquad \mathbb{H}^{(0)}=\bm{w}_0^\dagger  \ {\rm diag}(\bm\varepsilon^{(0)}) \ \bm{w}_0,
\ee
where $\bm{w}$ (resp. $\bm{w}_0$) are the single-particle eigenvectors and $\bm{\varepsilon}$ (resp. $\bm\varepsilon_0$) eigenvalues of the post- (resp. pre-) quench Hamiltonian matrix. 
By definition, the initial two-point correlation matrix is 
\be
\mathbb{G}(0)=\left[\braket{\Psi_0| \Cc_i^\dagger \Cc_j|\Psi_0}\right]_{i,j=1}^{2L}=\bm{w}_0\  \Pi \ \bm{w}_0^\dagger
\ee
where we defined the projector onto negative single-particle eigenstates:
\be
\Pi={\rm diag} \begin{pmatrix} p_1 \\ \vdots \\ p_L \end{pmatrix} ,\qquad p_j=\Theta(-\varepsilon_j)
\ee 
and $\Theta$ denotes the Heaviside step-function. At this point, by expanding the initial two-point correlation matrix in the eigenbasis of the post-quench Hamiltonian \eqref{xx-matrix}
\be
\mathbb{G}(0)=\bm{w}\  (\bm{w}^\dagger\ \bm{w}_0) \ \Pi \  (\bm{w}^\dagger_0 \ \bm{w}) \ \bm{w}^\dagger
\ee
the time evolution is simply given by
\be
\mathbb{G}(t)=\bm{w}\ {\cal U}(t)  \ (\bm{w}^\dagger\ \bm{w}_0) \ \Pi \  (\bm{w}^\dagger_0 \ \bm{w})  \ {\cal U}(t)^\dagger \ \bm{w}^\dagger
\ee
with time-evolution generator 
\be
{\cal U}(t)= {\rm diag} \exp(-\I t \bm{\varepsilon}).
\ee
%%%%%%%%%%%%%%%%%%%%%%%%%%%%%%%%%%%
One then determines the eigenvalues $\xi_j$ of the correlation matrix $\mathbb{G}(t)$ restricted to the subsystem $A$ to obtain the entanglement entropy according to Eq.~\eqref{EE-latt}.

\section{An alternative parametrization of the initial Fermi contour}\label{app:param-Jerome}

An alternative parametrization of the initial Fermi contour can be obtained by inverting Eq.~\eqref{kF-initial} in terms of the initial position
\be
x_0(k) =\frac{1}{\beta}\log\cos k, \qquad k\in[-\frac{\pi}{2},\frac{\pi}{2}]
\ee
and by subsequently introducing the real coordinate $s$ such that
\be
x_0(s)=-\frac{1}{\beta}\log\cosh(\beta s), \qquad k(s)=-2\arctan (\tanh \frac{\beta s}{2}).
\ee
This parametrization satisfies
\be
\frac{\dd x_0(s)}{\dd s}\equiv \sin k(s)= -\tanh(\beta s)
\ee
and therefore it ensures that the two-point correlation function along the initial Fermi contour is given simply as
\be\label{2pt-jerome}
\braket{\hat\Phi^+_\alpha(s_1)\hat{{\Phi}}^-_\alpha(s_2)}= (s_1-s_2)^{-2h_\alpha}.
\ee
At times $\tau>0$, the coordinate $s$ satisfies by construction the equation of motion 
\be\label{eq-of-motion-Jerome}
x=-\frac{1}{\beta}\log\cosh(\beta s) - \tau \tanh(\beta s)
\ee
which plays the same role as Eq.~\eqref{self-const}. Indeed, the latter is solved numerically for the Fermi point positions $s_{i=1,2}(x,\tau)$. Finally, a further conformal transformation
\be
s \mapsto w(s)= \exp(\I \pi s/L),
\ee
which modifies the two-point function in Eq.~\eqref{2pt-jerome} as
\be
\braket{\hat\Phi^+_\alpha(s_1)\hat{{\Phi}}^-_\alpha(s_2)}=\left(\frac{L}{\pi}\sin\frac{\pi(s_1-s_2)}{2L}\right)^{-2h_\alpha} ,
\ee
allows us to account for the finite-size of the chain. The entanglement profile is then given by
\be\label{S-Jerome}\begin{split}
S_1(x,\tau>0)&=\frac{1}{6} \log\left(\frac{L}{\pi} \left|\frac{\dd x}{\dd s}\right|_{s=s_1}^{1/2}\left|\frac{\dd x}{\dd s}\right|_{s=s_2}^{1/2} \left|\sin\frac{k(s_1)-k(s_2)}{2} \right|\left|\sin\frac{\pi(s_1-s_2)}{2L}\right|\right)+\Upsilon\\[4pt]
&=\frac{1}{6}\log \Big({\frac{L}{\pi}}\sqrt{\frac{\beta \tau}{\cosh^2(\beta s_1)}+\tanh(\beta s_1)}\sqrt{\frac{\beta \tau}{\cosh^2(\beta s_2)}+\tanh(\beta s_2)} \left|\sin\frac{\pi(s_1-s_2)}{2L}\right|\\[4pt]
& \times \frac{|\tanh\frac{\beta s_1}{2}- \tanh\frac{\beta s_2}{2}|}{\sqrt{1+\tanh^2\frac{\beta s_1}{2}}\sqrt{1+\tanh^2\frac{\beta s_2}{2}}}\Big) +\Upsilon,
\end{split}\ee
where we properly considered the cutoff dependence of Eq.~\eqref{cutoff}. Using Eq.~\eqref{eq-of-motion-Jerome} in \eqref{S-Jerome} and after some algebra, one can verify that Eq.~\eqref{S-Jerome} reduces to Eq.~\eqref{S-finite-beta}.

\section*{References}


\begin{thebibliography}{100}

\bibitem{Odor2004} G. \'Odor,
Universality classes in nonequilibrium lattice systems,
\href{https://journals.aps.org/rmp/abstract/10.1103/RevModPhys.76.663}{Rev. Mod. Phys. {\bf76}, 663 (2004)}

\bibitem{Dziarmaga2010} J. Dziarmaga,
Dynamics of a quantum phase transition and relaxation to a steady state,
\href{https://doi.org/10.1080/00018732.2010.514702}{Adv. Phys. {\bf59}, 1063 (2010)}.

\bibitem{pem-16} P. Calabrese, F. H. L. Essler, and G. Mussardo, Quantum Integrability in Out of Equilibrium Systems,
\href{http://iopscience.iop.org/article/10.1088/1742-5468/2016/06/064001}{J. Stat. Mech. (2016) 064001}.

\bibitem{Polkovnikov2011} A. Polkovnikov, K. Sengupta, A. Silva and M. Vengalattore,
Colloquium: Nonequilibrium dynamics of closed interacting quantum systems,
\href{https://journals.aps.org/rmp/abstract/10.1103/RevModPhys.83.863}{Rev. Mod. Phys. {\bf83}, 863 (2011)}

\bibitem{Bertini2021} B. Bertini, F. Heidrich-Meisner, C. Karrasch, T. Prosen, R. Steinigeweg and M. Znidaric,
Finite-temperature transport in one-dimensional quantum lattice models,
\href{https://journals.aps.org/rmp/abstract/10.1103/RevModPhys.93.025003}{Rev. Mod. Phys. {\bf93}, 025003 (2021)}.

\bibitem{Rigol2007} M. Rigol, V. Dunjko, V. Yurovsky, and M. Olshanii,
 Relaxation in a completely integrable many-body Quantum system: An Ab initio study of the dynamics of the highly excited states of 1D lattice hard-core bosons,
\href{https://journals.aps.org/prl/abstract/10.1103/PhysRevLett.98.050405}{Phys. Rev. Lett. {\bf98}, 050405 (2007)}

\bibitem{Rigol2008} M. Rigol, V. Dunjko, and M. Olshanii,
Thermalization and its mechanism for generic isolated quantum systems, 
\href{https://www.nature.com/articles/nature06838}{Nature {\bf 452}, 854 (2008)}

\bibitem{vr-16} L.~Vidmar and M.~Rigol, 
Generalized Gibbs ensemble in integrable lattice models, 
\href{http://dx.doi.org/10.1088/1742-5468/2016/06/064007}{J. Stat. Mech. (2016) 064007}.

%%%%%%%%%%%%%%%%%%%%%%%%%%%%%%%%%%

\bibitem{Bertini2016} B. Bertini, M. Collura, J. De Nardis and M. Fagotti,
 Transport in out-of-equilibrium XXZ chains: Exact profiles of charges and currents, 
\href{https://journals.aps.org/prl/abstract/10.1103/PhysRevLett.117.207201}{Phys. Rev. Lett. {\bf117}, 207201(2016)}.

\bibitem{Castro-Alvaredo2016} O. A. Castro-Alvaredo, B. Doyon and T. Yoshimura,
Emergent hydrodynamics in integrable quantum systems out of equilibrium,
\href{https://journals.aps.org/prx/abstract/10.1103/PhysRevX.6.041065}{Phys. Rev. X {\bf6}, 041065 (2016)}

\bibitem{Piroli2017} L. Piroli, J. De Nardis, M. Collura, B. Bertini and M. Fagotti,
 Transport in out-of-equilibrium XXZ chains: Nonballistic behavior and correlation functions,
\href{https://journals.aps.org/prb/abstract/10.1103/PhysRevB.96.115124}{Phys. Rev. B {\bf 96}, 115124 (2017)}

\bibitem{Bulchandani2017} V. B. Bulchandani, R. Vasseur, C. Karrasch and J. E. Moore,
 Solvable Hydrodynamics of Quantum Integrable Systems, 
\href{https://journals.aps.org/prl/abstract/10.1103/PhysRevLett.119.220604}{Phys. Rev. Lett. {\bf 119}, 220604 (2017)}

\bibitem{DeLuca2017} M. Collura, A. De Luca and J. Viti,
 Analytic solution of the domain-wall nonequilibrium stationary states, 
\href{https://journals.aps.org/prb/abstract/10.1103/PhysRevB.97.081111}{Phys. Rev. B {\bf 97}, 081111(R) (2017)}

\bibitem{Doyon2017} B. Doyon, J. Dubail, R. Konik and T. Yoshimura,
 Large-Scale Description of Interacting One-Dimensional Bose Gases: Generalized Hydrodynamics Supersedes Conventional Hydrodynamics,
\href{https://journals.aps.org/prl/abstract/10.1103/PhysRevLett.119.195301}{Phys. Rev. Lett. {\bf119}, 195301 (2017)}

\bibitem{DeLuca2017b} A. De Luca, M. Collura and J. De Nardis,
 Nonequilibrium spin transport in integrable spin chains: Persistent currents and emergence of magnetic domains, 
\href{https://journals.aps.org/prb/abstract/10.1103/PhysRevB.96.020403}{Phys. Rev. B {\bf 96}, 020403 (2017)}

\bibitem{Bulchandani2018} V. B. Bulchandani, R. Vasseur, C. Karrasch, and J. E. Moore,
 Bethe-Boltzmann hydrodynamics and spin transport in the XXZ chain, 
\href{https://journals.aps.org/prb/abstract/10.1103/PhysRevB.97.045407}{Phys. Rev. B {\bf97}, 045407 (2018)}

\bibitem{Doyon2018} B. Doyon, T. Yoshimura, and J. S. Caux,
 Soliton Gases and Generalized Hydrodynamics,
\href{https://journals.aps.org/prl/abstract/10.1103/PhysRevLett.120.045301}{Phys. Rev. Lett. {\bf120}, 045301 (2018)}

\bibitem{Bastianello2019} A. Bastianello, V. Alba, and J. S. Caux
Generalized Hydrodynamics with Space- Time Inhomogeneous Interactions, 
\href{https://journals.aps.org/prl/abstract/10.1103/PhysRevLett.123.130602}{Phys. Rev. Lett. {\bf123}, 130602 (2019)}

\bibitem{Mestyan2019} M. Mestyán, B. Bertini, L. Piroli and P. Calabrese,
Spin-charge separation effects in the low-temperature transport of one-dimensional Fermi gases,
\href{https://journals.aps.org/prb/abstract/10.1103/PhysRevB.99.014305}{Phys. Rev. B {\bf99}, 014305 (2019)}

\bibitem{Schemmer2019} M. Schemmer, I. Bouchoule, B. Doyon, and J. Dubail,
 Generalized Hydrodynamics on an Atom Chip,
\href{https://journals.aps.org/prl/abstract/10.1103/PhysRevLett.122.090601}{Phys. Rev. Lett.{\bf 122}, 090601 (2019)}

\bibitem{Bastianello2020a} A. Bastianello, J. De Nardis, and A. De Luca,
 Generalized hydrodynamics with dephasing noise,
\href{https://journals.aps.org/prb/abstract/10.1103/PhysRevB.102.161110}{Phys. Rev. B {\bf102}, 161110 (2020)}

\bibitem{Ruggiero2020} P. Ruggiero, P. Calabrese, B. Doyon and J. Dubail,
 Quantum Generalized Hydrodynamics,
\href{https://link.aps.org/doi/10.1103/PhysRevLett.124.140603}{Phys. Rev. Lett. {\bf124}, 140603 (2020)}

\bibitem{Malvania2020} N. Malvania, Y. Zhang, Y. Le, J. Dubail, M. Rigol, and D. S. Weiss,
Generalized hydrodynamics in strongly interacting 1D Bose gases,
 \href{https://arxiv.org/abs/2009.06651}{preprint - arXiv:2009.06651 (2020)}

\bibitem{Bouchoule2020a} I. Bouchoule and J. Dubail,
 Breakdown of Tan’s relation in lossy one-dimensional Bose gases,
\href{https://journals.aps.org/prl/abstract/10.1103/PhysRevLett.126.160603}{Phys. Rev. Lett. {\bf126}, 160603 (2021)}

\bibitem{Perfetto2020} G. Perfetto and B. Doyon,
Euler-scale dynamical fluctuations in non-equilibrium interacting integrable systems, 
\href{https://arxiv.org/abs/2012.06496}{preprint - arXiv:2012.06496}

\bibitem{Moller2020} F. S. M{\o}ller, G. Perfetto, B. Doyon and J. Schmiedmayer,
Euler-scale dynamical correlations in integrable systems with fluid motion, 
\href{https://scipost.org/SciPostPhysCore.3.2.016}{SciPost Phys. Core {\bf3}, 16 (2020)}

\bibitem{d-ls} B. Doyon, {\it Lecture notes on Generalised Hydrodynamics}, Les Houches Lecture notes,
\href{https://doi.org/10.21468/SciPostPhysLectNotes.18}{SciPost Phys. Lect. Notes {\bf 18} (2020)}.

\bibitem{Ilievski2017} E. Ilievski and J. De Nardis,
 Microscopic Origin of Ideal Conductivity in Integrable Quantum Models, 
\href{https://journals.aps.org/prl/abstract/10.1103/PhysRevLett.119.020602}{Phys. Rev. Lett. {\bf119}, 020602 (2017)}

\bibitem{Alba2021} V. Alba, B. Bertini, M. Fagotti, L. Piroli and P. Ruggiero,
Generalized-Hydrodynamic approach to Inhomogeneous Quenches: Correlations, Entanglement and Quantum Effects,
\href{http://arxiv.org/abs/2104.00656}{preprint - arXiv:2104.00656 (2021)}
%%%%%%%%%%%%%%%%%%%%%%%%%%%%%%%%%%%%%%

\bibitem{sc-08} S. Sotiriadis and J. Cardy, 
{Inhomogeneous Quantum Quenches},
\href{http://dx.doi.org/10.1088/1742-5468/2008/11/P11003}{J. Stat. Mech. (2008) P11003}

\bibitem{Dubail2017} J. Dubail, J.-M. St\'ephan, J. Viti and P. Calabrese,
 Conformal field theory for inhomogeneous one-dimensional quantum systems: the example of non-interacting Fermi gases,
\href{https://scipost.org/SciPostPhys.2.1.002}{SciPost Phys. {\bf2}, 2 (2017)}

\bibitem{Brun2017} Y. Brun and J. Dubail,
 One-particle density matrix of trapped one-dimensional impenetrable bosons from conformal invariance,
\href{https://scipost.org/SciPostPhys.2.2.012/pdf}{SciPost Phys. {\bf2}, 012 (2017)}

\bibitem{Brun2018} Y. Brun and J. Dubail,
 The Inhomogeneous Gaussian Free Field, with application to ground state correlations of trapped 1d Bose gases, 
\href{https://www.scipost.org/SciPostPhys.4.6.037/pdf}{SciPost Phys. {\bf4}, 037 (2018)}

\bibitem{Langmann2019} E. Langmann and P. Moosavi,
Diffusive Heat Waves in Random Conformal Field Theory,
\href{https://journals.aps.org/prl/abstract/10.1103/PhysRevLett.122.020201}{Phys. Rev. Lett. {\bf122}, 020201 (2019)}

\bibitem{Moosavi2019} P. Moosavi,
Inhomogeneous conformal field theory out of equilibrium,
\href{https://arxiv.org/abs/1912.04821}{preprint - arXiv:1912.04821 (2019)}

\bibitem{Ruggiero2019} P. Ruggiero, Y. Brun, and J. Dubail,
 Conformal field theory on top of a breathing one-dimensional gas of hard core bosons,
\href{https://www.scipost.org/SciPostPhys.6.4.051/pdf}{SciPost Phys. {\bf 6}, 051 (2019)}

\bibitem{Scopa2020} S. Scopa, L. Piroli and P. Calabrese,
 One-particle density matrix of a trapped Lieb Liniger anyonic gas, 
\href{https://iopscience.iop.org/article/10.1088/1742-5468/abaed1}{J. Stat. Mech. (2020) 093103}

\bibitem{Collura2020} M. Collura, A. De Luca, P. Calabrese, and J. Dubail,
 Domain wall melting in the spin-1/2 XXZ spin chain: Emergent Luttinger liquid with a fractal quasiparticle charge,
\href{https://journals.aps.org/prb/abstract/10.1103/PhysRevB.102.180409}{Phys. Rev. B {\bf 102}, 180409 (2020)}

\bibitem{Gluza2020} M. Gluza, P. Moosavi and S. Sotiriadis,
Breaking of Huygens-Fresnel principle in inhomogeneous Tomonaga-Luttinger liquids,
\href{https://arxiv.org/abs/2104.07751}{preprint - arXiv:2104.07751 (2021)}

\bibitem{Dean2019} D. S. Dean, P. Le Doussal, S. N. Majumdar, and G. Schehr,
 Noninteracting fermions in a trap and random matrix theory, 
\href{https://iopscience.iop.org/article/10.1088/1751-8121/ab098d}{J. Phys. A {\bf 52}, 144006 (2019)}

\bibitem{Dean2018} D. S. Dean, P. Le Doussal, S. N. Majumdar and G. Schehr,
 Wigner function of noninteracting trapped fermions, 
\href{https://journals.aps.org/pra/abstract/10.1103/PhysRevA.97.063614}{Phys. Rev. A {\bf97}, 063614 (2018)}

\bibitem{Smith2020} N. R. Smith, P. Le Doussal, S. N. Majumdar and G. Schehr,
Counting statistics for noninteracting fermions in a d-dimensional potential,
\href{https://journals.aps.org/pre/abstract/10.1103/PhysRevE.103.L030105}{Phys. Rev. E {\bf103}, L030105 (2021)}

\bibitem{Gautie2021} T. Gauti\'e, J.-P. Bouchaud and P. Le Doussal,
. Matrix kesten recursion, inverse-wishart ensemble and fermions in a morse potential, 
\href{https://arxiv.org/abs/2101.08082}{preprint - arXiv:2101.08082 (2021)}

\bibitem{DeBruyne2021} B. De Bruyne, D. S. Dean, P. Le Doussal, S. N. Majumdar and G. Schehr,
. Wigner function for noninteracting fermions in hard wall potentials, 
\href{https://arxiv.org/abs/2104.05068}{preprint - arXiv:2104.05068 (2021)}

\bibitem{Calabrese2006} P. Calabrese and J. Cardy,
 Time dependence of correlation functions following a quantum quench,
\href{https://journals.aps.org/prl/abstract/10.1103/PhysRevLett.96.136801}{Phys. Rev. Lett. {\bf 96}, 136801 (2006)}

\bibitem{Calabrese2007a} P. Calabrese and J. Cardy,
 Quantum quenches in extended systems, 
\href{https://iopscience.iop.org/article/10.1088/1742-5468/2007/06/P06008}{J. Stat. Mech. (2007) P06008}

\bibitem{Caux2013} J. S. Caux and F. H.L. Essler,
 Time evolution of local observables after quenching to an integrable model, 
\href{https://journals.aps.org/prl/abstract/10.1103/PhysRevLett.110.257203}{Phys. Rev. Lett. {\bf110}, 257203 (2013)}

\bibitem{Brockmann2014} M. Brockmann, B. Wouters, D. Fioretto, J. De Nardis, R. Vlijm and J. S. Caux,
Quench action approach for releasing the Néel state into the spin-1/2 XXZ chain, 
\href{https://iopscience.iop.org/article/10.1088/1742-5468/2014/12/P12009}{J. Stat. Mech. (2014)  P12009}

\bibitem{Ilievski2015} E. Ilievski, J. De Nardis, B. Wouters, J. S. Caux, F. H.L. Essler and T. Prosen,
 Complete Generalized Gibbs Ensembles in an Interacting Theory, 
\href{https://journals.aps.org/prl/abstract/10.1103/PhysRevLett.115.157201}{Phys. Rev. Lett. {\bf115}, 157201 (2015)}

\bibitem{Caux2016} J. S. Caux,
 The Quench Action, 
\href{https://iopscience.iop.org/article/10.1088/1742-5468/2016/06/064006}{J. Stat. Mech. (2016) 064006}

\bibitem{Calabrese2005} P. Calabrese and J. Cardy,
 Evolution of entanglement entropy in one-dimensional systems,
\href{https://iopscience.iop.org/article/10.1088/1742-5468/2005/04/P04010}{J. Stat. Mech. (2005) 04010}

\bibitem{Alba2017} V. Alba and P. Calabrese,
 Entanglement and thermodynamics after a quantum quench in integrable systems, 
\href{https://www.pnas.org/content/114/30/7947https://www.pnas.org/content/114/30/7947.abstract}{PNAS {\bf114}, 7947 (2017)}

\bibitem{ac-18} V. Alba and P. Calabrese, Entanglement dynamics after quantum quenches in generic integrable systems, 
\href{https://scipost.org/10.21468/SciPostPhys.4.3.017}{SciPost Phys. {\bf 4}, 017 (2018)}.

\bibitem{c-20} P. Calabrese, {\it Entanglement spreading in non-equilibrium integrable systems}, 
Lectures for Les Houches Summer School on ``Integrability in Atomic and Condensed Matter Physics", 
\href{https://doi.org/10.21468/SciPostPhysLectNotes.20}{SciPost Phys. Lect. Notes 20 (2020)}.

\bibitem{bfpc-18} B.~Bertini, M.~Fagotti, L.~Piroli, and P.~Calabrese, {Entanglement evolution and generalised hydrodynamics: noninteracting systems}, 
\href{https://doi.org/10.1088/1751-8121/aad82e}{J. Phys. A  {\bf 51}, 39LT01 (2018)}.

\bibitem{Alba2019} V. Alba, B. Bertini and M. Fagotti,
 Entanglement evolution and generalised hydrodynamics: Interacting integrable systems,
\href{https://scipost.org/SciPostPhys.7.1.005}{SciPost Phys. {\bf7}, 005 (2019)}

\bibitem{a-19} V. Alba, {Towards a generalized hydrodynamics description of Renyi entropies in integrable systems},
\href{https://doi.org/https://doi.org/10.1103/PhysRevB.99.045150}{Phys. Rev. B {\bf 99}, 045150 (2019)}.

\bibitem{Calabrese2007} P. Calabrese and J. Cardy,
 Entanglement and correlation functions following a local quench: A conformal field theory approach,
\href{https://iopscience.iop.org/article/10.1088/1742-5468/2007/10/P10004}{J. Stat. Mech. (2007) P10004}

\bibitem{Eisler2008} V. Eisler, D. Karevski, T. Platini and I. Peschel,
 Entanglement evolution after connecting finite to infinite quantum chains,
\href{https://iopscience.iop.org/article/10.1088/1742-5468/2008/01/P01023}{J. Stat. Mech. (2008) P01023}

\bibitem{Eisler2009} V. Eisler, F. Igl\'oi and I. Peschel,
 Entanglement in spin chains with gradients,
\href{https://iopscience.iop.org/article/10.1088/1742-5468/2009/02/P02011}{J. Stat. Mech (2009) P02011}

\bibitem{Igloi2009} F. Igl\'oi, Z. Szatm\'ari and Y. C. Lin,
 Entanglement entropy with localized and extended interface defects,
\href{https://journals.aps.org/prb/abstract/10.1103/PhysRevB.80.024405}{Phys. Rev. B {\bf 80}, 024405 (2009)}

\bibitem{Stephan2011} J.-M.  St\'ephan and J. Dubail,
 Local quantum quenches in critical one-dimensional systems: Entanglement, the Loschmidt echo, and light-cone effects,
\href{https://iopscience.iop.org/article/10.1088/1742-5468/2011/08/P08019}{J. Stat. Mech. (2011) P08019}

\bibitem{cc-13} M. Collura and P. Calabrese, {Entanglement evolution across defects in critical anisotropic Heisenberg chains},
\href{http://dx.doi.org/10.1088/1751-8113/46/17/175001}{J. Phys. A {\bf 46}, 175001 (2013)}

\bibitem{Calabrese2016} P. Calabrese and J. Cardy,
 Quantum quenches in 1 + 1 dimensional conformal field theories,
\href{https://iopscience.iop.org/article/10.1088/1742-5468/2016/06/064003}{J. Stat. Mech. (2016) 064003}

\bibitem{Antal1999} T. Antal, Z. R\'acz, A. R\'akos and G. M. Sch\"utz,
 Transport in the XX chain at zero temperature: Emergence of flat magnetization profiles,
\href{https://journals.aps.org/pre/abstract/10.1103/PhysRevE.59.4912}{Phys. Rev. E {\bf59}, 4912 (1999)}

\bibitem{Karevski2002} D. Karevski,
 Scaling behaviour of the relaxation in quantum chains,
\href{https://link.springer.com/article/10.1140/epjb/e20020139}{Eur. Phys. J. B {\bf27}, 147 (2002)}

\bibitem{Vicari2012} E. Vicari,
 Quantum dynamics and entanglement in one-dimensional Fermi gases released from a trap.
\href{https://journals.aps.org/pra/abstract/10.1103/PhysRevA.85.062324}{Phys. Rev. A {\bf85}, 062324 (2012)}

\bibitem{Alba2014} V. Alba and F. Heidrich-Meisner,
 Entanglement spreading after a geometric quench in quantum spin chains,
\href{https://journals.aps.org/prb/pdf/10.1103/PhysRevB.90.075144}{Phys. Rev. B {\bf90}, 075144 (2014)}

%%%%%%%%%%%%%%%%%%%%%%%%%%555

\bibitem{chl-08} P. Calabrese, C. Hagendorf, and P. Le Doussal,
{Time evolution of 1D gapless models from a domain-wall initial state: SLE continued?},
\href{http://dx.doi.org/10.1088/1742-5468/2008/07/P07013}{J. Stat. Mech. P07013 (2008)}

\bibitem{Allegra2016} N. Allegra, J. Dubail, J.-M. St\'ephan and J. Viti,
 Inhomogeneous field theory inside the arctic circle,
\href{https://iopscience.iop.org/article/10.1088/1742-5468/2016/05/053108}{J. Stat. Mech. (2016) 053108}

\bibitem{Bertini2018} B. Bertini, L. Piroli and P. Calabrese,
Universal Broadening of the Light Cone in Low-Temperature Transport,
\href{https://journals.aps.org/prl/abstract/10.1103/PhysRevLett.120.176801}{Phys. Rev. Lett. {\bf120}, 176801 (2018)}

\bibitem{Gruber2019} M. Gruber and V. Eisler,
Magnetization and entanglement after a geometric quench in the XXZ chain,
\href{https://journals.aps.org/prb/abstract/10.1103/PhysRevB.99.174403}{Phys. Rev. B {\bf99}, 174403 (2019)}

\bibitem{Lancaster2010} J. Lancaster and A. Mitra,
Quantum quenches in an XXZ spin chain from a spatially inhomogeneous initial state,
\href{https://journals.aps.org/pre/pdf/10.1103/PhysRevE.81.061134}{Phys. Rev. E {\bf 81}, 061134 (2010)}

\bibitem{mpc-10} J. Mossel, G. Palacios, J.-S. Caux, Geometric quenches in quantum integrable systems,
\href{http://dx.doi.org/10.1088/1742-5468/2010/09/L09001}{J. Stat. Mech. (2010) L09001}.

\bibitem{Antal2008} T. Antal, P. L. Krapivsky, and A. R\'akos,
 Logarithmic current fluctuations in nonequilibrium quantum spin chains,
\href{https://journals.aps.org/pre/abstract/10.1103/PhysRevE.78.061115}{Phys. Rev. E {\bf78}, 061115 (2008)}

\bibitem{Jin2021} T. Jin, T. Gauti\'e, A. Krajenbrink, P. Ruggiero and T. Yoshimura,
Interplay between transport and quantum coherences in free fermionic systems,
\href{https://arxiv.org/abs/2103.13371}{preprint - arXiv:2103.13371 (2021)}

\bibitem{Cazalilla2004} M. A. Cazalilla,
 Bosonizing one-dimensional cold atomic gases,
\href{https://iopscience.iop.org/article/10.1088/0953-4075/37/7/051}{J. Phys. B {\bf 37}, 1 (2004)}

\bibitem{Giamarchi2007} T. Giamarchi,
\href{https://oxford.universitypressscholarship.com/view/10.1093/acprof:oso/9780198525004.001.0001/acprof-9780198525004}{{\it Quantum Physics in One Dimension}, Oxford Univ. Press (2007)}

\bibitem{next-pub} P. Calabrese, J. Dubail, A. Krajenbrink, and S. Scopa, {\it in preparation}

\bibitem{Jordan1928} P. Jordan and E. Wigner, \"Uber das Paulische \"Aquivalenzverbot. 
\href{https://link.springer.com/article/10.1007/BF01331938}{Z. Phys. {\bf 47}, 631 (1928)}.

\bibitem{Jin2004} B. Q. Jin and V. E. Korepin,
 Quantum spin chain, Toeplitz determinants and the Fisher-Hartwig conjecture,
\href{https://link.springer.com/article/10.1023/B:JOSS.0000037230.37166.42}{J. Stat. Phys. {\bf116}, 79 (2004)}

\bibitem{fg-21} S. Fraenkel and M. Goldstein, Entanglement Measures in a Nonequilibrium Steady State: Exact Results in One Dimension,
 \href{https://arxiv.org/abs/2105.00740 }{preprint - arXiv:2105.00740  (2021)}.

\bibitem{Wendenbaum2013} P. Wendenbaum, M. Collura, and D. Karevski,
 Hydrodynamic description of hard-core bosons on a Galileo ramp,
\href{https://journals.aps.org/pra/abstract/10.1103/PhysRevA.87.023624}{Phys. Rev. A {\bf87}, 023624 (2013)}

\bibitem{Bastianello2020} A. Bastianello, J. Dubail, and J.-M. St\'ephan,
 Entanglement entropies of inhomogeneous Luttinger liquids,
\href{https://doi.org/10.1088/1751-8121/ab7580}{J. Phys. A {\bf 53}, 23 (2020)}

\bibitem{Eisler2013} V. Eisler and Z. R\'acz,
Full counting statistics in a propagating quantum front and random matrix spectra,
\href{https://link.aps.org/doi/10.1103/PhysRevLett.110.060602}{Phys. Rev. Lett. {\bf110}, 060602 (2013)}

\bibitem{Moriya2019} H. Moriya, R. Nagao and T. Sasamoto,
Exact large deviation function of spin current for the one dimensional XX spin chain with domain wall initial condition,
\href{https://iopscience.iop.org/article/10.1088/1742-5468/ab1dd6}{J. Stat. Mech. (2019) 063105}

\bibitem{Bettelheim2011} E. Bettelheim and P. B. Wiegmann,
Universal Fermi distribution of semiclassical nonequilibrium Fermi states,
\href{https://journals.aps.org/prb/abstract/10.1103/PhysRevB.84.085102}{Phys. Rev. B {\bf84}, 085102 (2011)}

\bibitem{Bettelheim2012} E. Bettelheim and L. Glazman,
Quantum Ripples Over a Semiclassical Shock,
\href{https://journals.aps.org/prl/abstract/10.1103/PhysRevLett.109.260602}{Phys. Rev. Lett. {\bf109}, 260602 (2012)}

\bibitem{Wigner1997} E. P. Wigner,
{On the quantum correction for thermodynamic equilibrium},
\href{https://link.springer.com/chapter/10.1007/978-3-642-59033-7_9}{Springer-Verlag Berlin Heidelberg (1997)}

\bibitem{Hinarejos2012} M. Hinarejos, A. P\'erez and M. C. Banuls,
 Wigner function for a particle in an infinite lattice,
\href{https://iopscience.iop.org/article/10.1088/1367-2630/14/10/103009/meta}{New J. Phys. {\bf 14}, 103009 (2012)}

\bibitem{Fagotti2017} M. Fagotti,
Higher-order generalized hydrodynamics in one dimension: The noninteracting test,
\href{https://journals.aps.org/prb/abstract/10.1103/PhysRevB.96.220302}{Phys. Rev. B {\bf96}, 220302 (2017)}

\bibitem{Fagotti2020} M. Fagotti,
Locally quasi-stationary states in noninteracting spin chains,
\href{https://scipost.org/10.21468/SciPostPhys.8.3.048}{SciPost Phys. {\bf8}, 048 (2020)}

\bibitem{Moyal1949} J. E. Moyal,
 Quantum mechanics as a statistical theory,
\href{https://www.cambridge.org/core/journals/mathematical-proceedings-of-the-cambridge-philosophical-society/article/abs/quantum-mechanics-as-a-statistical-theory/9D0DC7453AD14DB641CF8D477B3C72A2}{Math. Proc. Cambridge Philos. Soc. {\bf45}, 99 (1949)}

\bibitem{Calabrese2004} P. Calabrese and J. Cardy
Entanglement entropy and quantum field theory,
\href{https://iopscience.iop.org/article/10.1088/1742-5468/2004/06/P06002}{J. Stat. Mech. (2004) P06002}

\bibitem{Cardy2008} J. L. Cardy, O. A. Castro-Alvaredo and B. Doyon,
 Form factors of branch-point twist fields in quantum integrable models and entanglement entropy,
\href{https://link.springer.com/article/10.1007/s10955-007-9422-x}{J. Stat. Phys. {\bf130}, 129 (2008)}

\bibitem{Calabrese2009} P. Calabrese and J. Cardy,
 Entanglement entropy and conformal field theory,
\href{https://iopscience.iop.org/article/10.1088/1751-8113/42/50/504005}{J. Phys. A {\bf 42}, 504005 (2009)}

\bibitem{Calabrese2010} P. Calabrese and F. H.L. Essler,
 Universal corrections to scaling for block entanglement in spin-1/2 XX chains,
\href{https://iopscience.iop.org/article/10.1088/1742-5468/2010/08/P08029}{J. Stat. Mech. (2010) P08029}

\bibitem{cmv-11} P. Calabrese, M. Mintchev, and E. Vicari, {Entanglement Entropy of One-Dimensional Gases}, 
\href{https://doi.org/10.1103/PhysRevLett.107.020601}{Phys. Rev. Lett. {\bf 107}, 020601 (2011)}.

\bibitem{cmv-11b} P. Calabrese, M. Mintchev, and E. Vicari,  {The entanglement entropy of one-dimensional systems in continuous and homogeneous space}, 
\href{http://dx.doi.org/10.1088/1742-5468/2011/09/P09028}{J. Stat. Mech. P09028 (2011)}.

\bibitem{Dubessy2020} R. Dubessy, J. Polo, H. Perrin, A. Minguzzi and M. Olshanii,
Universal shock-wave propagation in one-dimensional Bose fluids, 
\href{https://journals.aps.org/prresearch/abstract/10.1103/PhysRevResearch.3.013098}{Phys. Rev. Res. {\bf3}, 013098 (2021)}

\bibitem{Lieb1961} E. H. Lieb, T. Schultz and D. Mattis, 
Two Soluble Models of an Antiferromagnetic Chain,
\href{https://reader.elsevier.com/reader/sd/pii/0003491661901154?token=0189AFCFF0917F2AD60912AE6437245300497126C5E6E21D5D06CFBFA3D035CC1441288ADCA5FCD47C9D89CEB12BFCD5&originRegion=eu-west-1&originCreation=20210727065759}{Ann. of Phys. {\bf 16}, 407-466 (1961)}

\bibitem{Mukherjee2007} V. Mukherjee, U. Divakaran, A. Dutta and D. Sen,
Quenching dynamics of a quantum XY spin-1/2 chain in a transverse field,
\href{https://journals.aps.org/prb/pdf/10.1103/PhysRevB.76.174303}{Phys. Rev. B {\bf 76}, 174303 (2007)}

\bibitem{Barmettler2010} P. Barmettler, M. Punk, V. Gritsev, E. Demler and E. Altman,
Quantum quenches in the anisotropic spin-1/2 Heisenberg chain: different approaches to many-body dynamics far from equilibrium,
\href{https://iopscience.iop.org/article/10.1088/1367-2630/12/5/055017/pdf}{New J. Phys. {\bf 12}, 055017 (2010)}

\bibitem{Yoshinaga2021} A. Yoshinaga, 
Ballistic propagation in the one-dimensional XY model,
\href{https://iopscience.iop.org/article/10.1088/1742-5468/abcd37}{J. Stat. Mech. (2021) 013103}

\bibitem{p-12} I. Peschel,  Entanglement in solvable many-particle models,
 \href{http://dx.doi.org/10.1007/s13538-012-0074-1}{Braz. J. Phys. {\bf 42}, 267 (2012)}.

\bibitem{Peschel1999a} I. Peschel, M. Kaulke and \"O. Legeza,
 Density-matrix spectra for integrable models,
\href{https://doi.org/10.1002/(SICI)1521-3889(199902)8:2<153::AID-ANDP153>3.0.CO;2-N}{Ann. Phys. {\bf8}, 153 (1999)}

\bibitem{Chung2001} M. C. Chung and I. Peschel,
 Density-matrix spectra of solvable fermionic systems,
\href{https://journals.aps.org/prb/abstract/10.1103/PhysRevB.64.064412}{Phys. Rev. B {\bf64}, 064412 (2001)}

\bibitem{Peschel2003} I. Peschel,
 Calculation of reduced density matrices from correlation functions,
\href{https://iopscience.iop.org/article/10.1088/0305-4470/36/14/101}{J. Phys. A {\bf36}, L205 (2003)}

\bibitem{Peschel2004} I. Peschel,
 On the reduced density matrix for a chain of free electrons,
\href{https://iopscience.iop.org/article/10.1088/1742-5468/2004/06/P06004}{J. Stat. Mech. (2004) P06004}

\bibitem{Peschel2009} I. Peschel and V. Eisler,
 Reduced density matrices and entanglement entropy in free lattice models,
\href{https://iopscience.iop.org/article/10.1088/1751-8113/42/50/504003}{J. Phys. A {\bf42}, 504003 (2009)}


\end{thebibliography}
\end{document}